\documentclass[a4paper,10pt]{article}
\usepackage{amsmath,amsfonts,amssymb,graphics,graphicx,epsfig,color,bbm,subfigure}

\newcommand{\R}{\mathbbm{R}}

\newcommand{\gr}[1]{\boldsymbol{#1}}
\newcommand{\be}{\begin{equation}}
\newcommand{\ee}{\end{equation}}
\newcommand{\bea}{\begin{eqnarray}}
\newcommand{\eea}{\end{eqnarray}}
\newcommand{\ket}[1]{|#1\rangle}
\newcommand{\bra}[1]{\langle#1|}

\newcommand{\N}{{\cal N}}
\newcommand{\eq}[1]{Eq.~(\ref{#1})}

\begin{document}
\title{Quantifying decoherence in continuous variable systems}
\author{A.\ Serafini,$^{1,2}$
M.\ G.\ A.\ Paris,$^{3}$ F.\ Illuminati,$^{1}$ and S.\ De Siena$^{1}$}
\date{\footnotesize
$^{1}$ Dipartimento di Fisica ``E. R. Caianiello'',
Universit\`a di Salerno, INFM UdR Salerno,\\
\footnotesize INFN Sezione Napoli, Gruppo Collegato Salerno,
Via S. Allende, 84081 Baronissi (SA), Italy\\ \vspace*{1mm}
\footnotesize$^{2}$ Department of Physics and Astronomy, 
University College of London,\\ 
\footnotesize Gower Street, London, WC1E 6BT, United Kingdom\\
\footnotesize$^{3}$ Dipartimento di Fisica and INFM, Universit\`a di Milano,
Milano, Italy}

\maketitle

%%%%%%%%%%%%%%%%%%%%%%%%%%%%%%%%%%%%%%%%%%%%%%%%%%
\begin{abstract}
We present a detailed report on the decoherence of quantum
states of continuous variable systems under the action
of a quantum optical master equation resulting from the 
interaction with general Gaussian uncorrelated environments. 
The rate of decoherence is quantified by relating it to the
decay rates of various, complementary measures of the quantum
nature of a state, such as the purity, some nonclassicality
indicators in phase space and, for two-mode states, 
entanglement measures and total correlations between the modes. 
Different sets of physically relevant initial configurations are 
considered, including one- and two-mode Gaussian states, 
number states, and coherent superpositions.
Our analysis shows that, generally, the use of initially
squeezed configurations does not help to preserve the coherence 
of Gaussian states, whereas it can be effective in protecting coherent 
superpositions of both number states and Gaussian wave packets.
\end{abstract}
%%%%%%%%%%%%%%%%%%%%%%%%%%%%%%%%%%%%%%%%%%%%%%%%%%

\section{Introduction}\label{intro}

Beyond their fundamental interest in the physics of elementary
particles (quantum electrodynamics and its standard-model generalizations),
in quantum optics, and in condensed matter theory, 
continuous variable systems are beginning to play an outstanding role 
in quantum communication and information theory \cite{pati,rmp}, as shown by the 
first spectacular implementations of deterministic teleportation schemes
and quantum key distribution protocols in quantum optical settings \cite{tele,gran}.

In all such practical instances the information contained in a given quantum state
of the system, so precious for the realization of any specific task, is
constantly threatened by the unavoidable interaction with the environment.
Such an interaction entangles the system of interest with 
the environment, causing any amount of information to be scattered and lost
in the (infinite) Hilbert space of the environment. It is important to remark
that this information is irreversibly lost, since the degrees of freedom of the
environment are out of the experimental control. The overall process,
corresponding to a non unitary evolution of the system, is commonly referred to as
decoherence \cite{leggett83,zurek91}. It is thus of crucial importance
to develop proper methods to quantify the rate of decoherence, both for
its understanding and for building optimal strategies to reduce and/or
suppress it.

In this work we study the decoherence of generic states of continuous 
variable systems whose evolution is ruled by optical master equations 
in general Gaussian uncorrelated environments.
The rate of decoherence is quantified
by analysing the evolution of global entropic measures, of nonclassical
indicators and, for two-mode states, of entanglement and correlations 
quantified by the mutual information and by the logarithmic negativity.
Several initial states of major interest are considered.

The plan of the paper is as follows. In Section \ref{nota} we
introduce the notation and define the systems of interest, together with
the quantities we will adopt to quantify decoherence. In Section \ref{master}
we introduce and solve the quantum optical master equation and its
corresponding phase space diffusive equations, discussing some general
properties of the nonunitary evolution. In Sections \ref{1mode}-\ref{2mode}
we provide a detailed study of the decoherence of single mode Gaussian states,
cat-like states, number states and two-mode Gaussian states. Finally,
in Section \ref{concl} we review and comment the relevant results.

\section{Notation and basic concepts}\label{nota}
The system we address is a canonical infinite dimensional system constituted by
a set of $n$ `modes'. Each mode $i$ is described by
a pair of canonical conjugate operators $\hat{x}_i$, $\hat{p}_i$ acting on a
denumerable Hilbert space ${\cal H}_i$. The space ${\cal H}_i$ is
spanned by a number basis $\{\ket{n}_k\}$ of eigenstates of the operator $\hat{n}_k\equiv
a_k^{\dag}a_k$, which represents the Hamiltonian of the non interacting mode. In terms
of the ladder operators $a_k$ and $a_k^{\dag}$ one has $\hat{x}_k=(a_k+a_k^{\dag})/\sqrt{2}$
and $\hat{p}_k=i(a_k^{\dag}-a_k)/\sqrt{2}$.
Let us group together the canonical operators
in the vector of operators $\hat{R}=(\hat{x}_1,\hat{p}_1,\ldots,\hat{x}_n,\hat{p}_n)$.
The canonical commutation relations regulate the commutation properties of the operators:
\[ [\hat{R}_k,\hat{R}_l]=i\Omega_{kl} \; ,\]
where $\Omega$ is the symplectic form
\be
\Omega=\bigoplus_{i=1}^{n}\omega\, , \quad \omega=
\left(\begin{array}{cc}
0&1\\
-1&0
\end{array}\right)\, . \label{symform}
\ee 
The canonical operators $\hat{R}_i$ may be second quantized bosonic field operators or 
position and momentum operators of a material harmonic oscillator.
The eigenstates of $a_i$ constitutes the important
set of coherent states, which is overcomplete in the Hilbert space ${\cal H}_i$.
Coherent states result from applying to the vacuum $\ket{0}$ the 
single-mode Weyl displacement operators
$D_i(\alpha)=\,{\rm e}^{i\alpha a_i^{\dag}-\alpha^{*}a_i}$: $\ket{\alpha}_i=
D_i(\alpha)\ket{0}$. 

The states of the system are the set of 
positive trace class operators $\{\varrho\}$ on the Hilbert space 
${\cal H}=\otimes_{i=1}^n {\cal H}_i$. However, the complete description of any 
quantum state $\varrho$ of such an infinite dimensional system can be provided by 
one of its $s$-ordered characteristic functions \cite{barnett}
\be
\chi_s (X) = \,{\rm Tr}\,[\varrho D_{X}] \,{\rm e}^{s\|X\|^2/2} \; ,
\ee
with $X\in\R^{2n}$, $\|\centerdot\|$ standing for the Euclidean norm $\R^{2n}$ 
and the $n$-mode Weyl operator defined as
\[
D_{X}=\,{\rm e}^{i\hat{R}^T \Omega X},\quad X\in \R^{2n} \, . 
\] 

The family of characteristic functions is in turn related, via complex Fourier transform, 
to the quasi-probability distributions $W_s$, which constitutes 
another set of complete descriptions of the quantum states
\be
W_s(X)=\frac{1}{\pi^2}\int_{{\mathbbm R}^{2n}}\,{\rm d}^{2n}K \chi_s(K)
\,{\rm e}^{iK^T \Omega X} \, . 
\ee
The vector $X$ belongs to the 
space $\Gamma=({\mathbbm R}^{2n},\Omega)$, which is called phase space
in analogy with classical Hamiltonian dynamics.
As well known, there exist states for which
the function $W_s$ is not a regular probability distribution for any $s$, because 
it can in general be singular or assume negative values. 
Note that the value $s=-1$ corresponds to the Husimi `Q-function' 
$W_{-1}(X)=\bra{X}\varrho\ket{X}/\pi$, $\ket{X}$ being a
tensor product of coherent states 
satisfying 
\be
a_i\ket{X}=\frac{X_{2i-1}+iX_{2i}}{\sqrt{2}}\ket{X}\quad \forall i=1,\ldots n \, ,
\label{ncohe}\ee 
and always yields a regular probability 
distribution. The case $s=0$ correponds to the so called Wigner function,
which will be denoted simply by $W$. Likewise, for the sake of simplicity, $\chi$ 
will stand for the simmetrically ordered characteristic function $\chi_0$.\par  

As a meausure of `nonclassicality' of the quantum state $\varrho$,
the quantity $\tau_{\varrho}$, referred to as `nonclassical depth', has been 
proposed in Ref.~\cite{lee} and subsequently employed by many authors
\be
\tau_{\varrho}=\frac{1-\bar{s}_{\varrho}}{2} \; ,
\ee
where $\bar{s}_{\varrho}$ is the supremum of the set of values $\{s\}$
for which the quasiprobability
function $W_s$ associated to the state $\varrho$ can be regarded as a (positive semidefinite
and non singular)
probability distribution.
We mention that a nonzero nonclassical depth
has been shown to be a prerequisite for the generation of continuous variable entanglement
\cite{kim02} and is strictly related to the efficiency of teleportation protocols \cite{takeoka}.
As one should expect, $\tau_{\ket{n}\bra{n}}=1$
for number states (which are actually the most deeply quantum and `less classical' ones),
whereas $\tau_{\ket{\alpha}\bra{\alpha}}=0$ for coherent states
(which are often referred to as `the most classical' among the quantum states).
We note that the nonclassical depth
can be interpreted as the minimum number of thermal photons
which has to be added to a quantum state in order to erase all the `quantum
features'
of the state.\footnote{This heuristic statement can be made more rigorous by assuming that a
given state owns `quantum features' if and only if its P-representation is more
singular than a delta function (which is the case for coherent states) \cite{lee}.}
While quite effective, the nonclassical depth is not always easily evaluated for
relevant quantum states (with the major exception of Gaussian states, see the following).

Therefore, it will be convenient to exploit also another indicator of nonclassicality,
more recently introduced \cite{benedict}. By virtue of intuition, one should expect that remarkable
non classical features should show up for quantum states whose Wigner functions
assume negative values. In fact, for such states, an equivalent interpretation in terms of
classical probabilities and correlations is denied.\footnote{This is the reason why in the search
of CV states able to violate Bell inequalities one is lead to consider states
with non positive Wigner functions.} These considerations have lead to the folllowing 
definition of the quantity $\xi$, which we will refer to as the `negative part' of the state 
$\varrho$
\be
\xi=\int \,{\rm d}^{2n}X |W(X)| - 1 \; ,  
\ee
which simply corresponds the doubled volume of the negative part of the Wigner function $W$
associated to $\varrho$ (the normalization of $W$ has been exploited). 
\par

This work will be partly focused on Gaussian states, defined as the states with Gaussian 
Wigner function or characteristic function $\chi$. 
Such states are completely characterized by first and second 
moments of the quadrature operators, respectively embodied by the first moment 
vector $\bar{X}$ and by the covariance matrix (CM) $\gr{\sigma}$, whose 
entries are, respectively
\bea
\bar{X}_i&\equiv& \langle \hat R_i \rangle \; , \\
\sigma_{ij}&\equiv &\frac{\langle \hat R_i \hat R_j + \hat R_j \hat R_i \rangle}{2}
- \langle\hat R_i\rangle\langle\hat R_j\rangle \, .
\eea
The covariance matrix of a physical state has to satisfy the following 
uncertainty relation, reflecting the positivity of the density matrix \cite{simon87}
\be
\gr{\sigma}+i \frac{\Omega}{2}\ge 0 \; . \label{unce}
\ee
The Wigner function of a Gaussian state can be written as
\be
W(X)=\frac{1}{\pi\sqrt{\,{\rm Det}\,\gr{\sigma}}}
\,{\rm e}^{-\frac12 (X-\bar{X})^T \gr{\sigma}^{-1} (X-\bar{X})} , 
\quad \xi\in\Gamma \, , \label{gauwig}
\ee
corresponding to the following characteristic function 
\be
\chi(X)=
\,{\rm e}^{-\frac12 (X-\bar{X})^T \gr{\sigma} (X-\bar{X})
+iX^T\Omega\bar{X}} , \, .\label{gauchar}
\ee

A tensor product of coherent states $\ket{\bar{X}}$
[simultaneous eigenstate of all
the $a_i$'s according to \eq{ncohe}] is a Gaussian state
with covariance matrix $\gr{\sigma}=\frac12{\mathbbm 1}$ and first moment
vector $\bar{X}$. In  phase space this amounts to simply displacing the Wigner function of
the vacuum.\par
A single mode of the radiation of frequency $\omega$
at thermal equilibrium at temperature $T$ is described 
by a Gaussian Wigner function as well. Its covariance matrix $\gr{\nu}$ is isotrope: 
$\gr{\nu}=\nu{\mathbbm 1}_2$ with $\nu=[\exp(\omega/T)+1]/[2\exp(\omega/T)-2]\ge 1/2$
(natural units are understood), while its first moments are null. 

The set of operations generated by second order 
polynomials in the quadrature operators are especially relevant in dealing with Gaussian states.
Such operations correspond to symplectic transformations in phase space, 
{\em i.e.~}to linear transformations preserving the symplectic form $\Omega$
\cite{prama}. 
Formally, a $2n\times 2n$ matrix $S$ correspond to a symplectic transformation 
(on a $n$-mode phase space) if and only if
\[
S^T \Omega S = \Omega \; .
 \]
Simplectic transformations act linearly on first moments and
by congruence on covariance matrices: $\gr{\sigma}\mapsto
S^T \gr{\sigma} S$. Ideal beam splitters and squeezers are described by simplectic
transformations. In fact single and two-mode squeezings are described
by the operators $S_{ij,r,\varphi}=\,{\rm e}^{\frac12 (\varepsilon a_i^{\dag}a_j^{\dag}
-\varepsilon^{*}a_ia_j)}$ with $\varepsilon=r\,{\rm e}^{i2\varphi}$,
resulting in single-mode squeezing of mode $i$ for $i=j$.
Beam splitters are described by the operator $O_{ij,\theta}=
\,{\rm e}^{\theta a_i^{\dag}a_j-\theta a_ia_j^{\dag}}$, corresponding 
to simplectic rotations in phase space.

A theorem by Williamson \cite{williamson} ensures that any $n$-mode CM $\gr{\sigma}$
can be written as 
\be
\gr{\sigma} = S^T \gr{\nu} S \; ,
\ee
where $S$ is a (non unique) simplectic transformation and 
\be
\gr{\nu}=\bigoplus_{i=1}^{n}\left(\begin{array}{cc}
\nu_i&0\\
0&\nu_i 
\end{array}\right) \, . 
\ee 
The Gaussian state with null first moments and CM $\gr{\nu}$
is a tensor product\footnote{as can be promptly seen from the definition 
of the characteristic functions, tensor products in Hilbert spaces correspond to 
direct sums in phase spaces} 
of thermal states with average photon numbers $\nu_i-1/2$ and density matrices
$\rho_{\nu_i}$ 
\bea
\rho_{\nu_i} = \frac{2}{2\nu_i+1}\sum_{k=0}^{\infty}\left(
\frac{\nu_i-\frac12}{\nu_i+\frac12}\right)^k \ket{k}\bra{k}  \, .
\eea 
The set $\{\nu_i\}$ is referred to as the simplectic spectrum of $\gr{\sigma}$,
the quantities $\nu_i$'s being the symplectic eigenvalues, which are just the 
eigenvalues of the matrix $|i\Omega\gr{\sigma}|$. The uncertainty
relation Ineq.~(\ref{unce}) can be simply written in terms of the symplectic 
eigenvalues
\be
\nu_i\ge \frac12 \quad \forall\, i=1,\ldots,n \, .
\label{sympunc}
\ee

As a last remark about Gaussian states, we briefly address their nonclassicality. 
Of course, for the negative part of a Gaussian state one has $\xi=0$. 
Remarkably, such an indicator does not detect squeezed states as non classical.
We point out that this fact is not detrimental to the indicator $\xi$. As a matter 
of facts any Gaussian state can be reproduced in classical stochastic systems described 
by probability distribution, where even an uncertainty relation analogous to 
Ineq.~(\ref{unce}) has to be introduced. On the other hand, the nonclassical 
depth of a $n$-mode Gaussian state $\varrho$ depends only on the smallest
(orthogonal, not symplectic) eigenvalue $u$ of the CM $\gr{\sigma}$, which  
is usually referred to as the `generalized squeeze variance' \cite{simon94}.
The indicator $\tau$ detects a Gaussian state as a nonclassical one 
(for which $\tau>0$) if a canonical quadrature 
(possibly resulting from the linear combination
of the quadratures of the separate modes) exists whose variance is below $1/2$.
The explicit expression for the nonclassical depth
of a Gaussian state $\varrho$ with CM $\gr{\sigma}$ reads
\be
\tau_{\varrho}=\max\left[\frac{1-2u}{2},0\right] \; . \label{ncgau}
\ee
As we have already remarked, coherent states have null nonclassical depth. One has 
to squeeze the covariances to achieve nonclassical features, like subpoissonian
photon number distributions. Regardless of the amount of squeezing, no Gaussian 
state can go beyond the threshold of $\tau_{\varrho}=1/2$.

In general, the degree of mixedness of a quantum state $\varrho$ of a system with a
$d$-dimensional Hilbert space
can be characterized
by means of the so called purity $\mu=\,{\rm Tr}\,\varrho^{2}$, taking the value 
$1$ on pure states (for which $\varrho^2=\varrho$) and going to $1/d$ 
(that is $0$ in 
infinite dimensional Hilbert spaces) for `maximally mixed' states. The purity is a simple function 
of the linear entropy $S_L = (1-\mu)d/(d-1)$ and of the Renyi `2-entropy' $S_2=-\ln\mu$, 
which is endowed with the agreeable feature of being additive on tensor product states. 
While other entropic measures, like the Von Neumann entropy, could have been taken 
into account, the purity has the remarkable advantage of being easily computable in terms 
of the Wigner function $W(X)$. Moreover, the global and marginal purities 
({\em i.e.~}the purities of the state of the whole system and of the reduced states 
of the subsystems) have been shown
to provide essential information about the quantum correlations of both two-mode 
Gaussian states \cite{adesso,adesso2} and multipartite, multimode Gaussian states 
\cite{adesso3,unitaryloc,adesso4}. We also remark that strategies have been proposed
to directly measure such a quantity, either by quantum networks \cite{ekefil} or 
by schemes based on single photon detections \cite{fiura}.

Exploiting the basic properties of the Wigner representation,
one has simply
\be
\mu=\pi\int W^2(X)\,{\rm d}^{2n}X=\frac{1}{2\pi}\int_{\R^{2n}}|\chi(X)|^2
\,{\rm d}^{2n}X  \; . \label{wigpur}
\ee
For Gaussian states this integral is straightforwardly evaluated, giving
\be
\mu=\frac{1}{2^n\sqrt{\,{\rm Det}\,\gr{\sigma}}} \; .\label{purgau}
\ee
The same result could have been achieved by exploiting Williamson theorem and the 
unitary invariance of $\mu$. This is indeed the way to compute general entropic 
measures of Gaussian states \cite{adesso2}. In particular, the von Neumann entropy
$S_V=-\,{\rm Tr}\,[\varrho\ln\varrho]$ of the   
Gaussian state $\varrho$ is easily expressed in terms of the $n$ symplectic  $\nu_i$'s of
the $2n\times 2n$ covariance matrix $\gr{\sigma}$ \cite{holevo99, serafozzi}
\be
S_V=\sum_{i=1}^{n}f(\nu_i) \; , \label{vneu}
\ee
with the bosonic entropic function $f(x)$ defined by
\[
f(x)=(x+\frac12)\ln(x+\frac12)-(x-\frac12)\ln(x-\frac12) \; .
\]
This formula will be useful in quantifying the total (quantum plus classical)
correlations between different modes in two-mode Gaussian states, which will be addressed
in the following. In general, the total correlations belonging to a bipartite quantum state 
$\varrho$ may be quantified by its mutual information $I$, defined as 
$I=S_V(\varrho_1)+S_V(\varrho_2)-S_V(\varrho)$, where $\varrho_i$ refers to 
the reduced state obtained by tracing over the variables of the party $j\neq i$
\cite{vedral01}. 

Finally, we introduce the definition of logarithmic negativity for bipartite quantum states, 
which will be exploited in the following in quantifying the entanglement ({\em i.e.}~the amount
of quantum correlations) of two-mode Gaussian states. 
For such states separability is equivalent to positivity of the partial transpose $\tilde{\varrho}$ 
(PPT criterion)\cite{simon00,duan00}.\footnote{The partial 
transpose $\tilde{\varrho}$ is obtained by the bipartite state 
$\varrho$ by transposing the Hilbert space of only one of the two parties.} 
The negativity ${\cal N}(\varrho)$ of the state $\varrho$ is defined as \cite{vidwer,eiserth}
\be
{\cal N}(\varrho)=\frac{\|\tilde{\varrho}\|_1-1}{2} \; ,
\ee
where $\|\hat{o}\|=\,{\rm Tr}\,\sqrt{\hat{o}^{\dag}\hat{o}}$ stands for the trace norm
of operator $\hat{o}$. 
The quantity ${\cal N}(\varrho)$, being the modulus of the sum 
of the negative eigenvalues of $\tilde{\varrho}$, quantifies the extent to which 
$\tilde{\varrho}$ fails to be positive. 
The logarithmic negativity $E_{\cal N}$ 
is then just defined as $E_{\cal N}=\ln \|\tilde{\varrho}\|_1$. From an operational
point of view, the logarithmic negativity constitutes an upper bound to the distillable 
entanglement \cite{vidwer} 
and is directly related to the entanglement cost under PPT preserving 
operations \cite{auden03}.

%%%%%%%%%%%%%%%%%%%%%%%%%%%%%%%%%%%%%%%%%%%%%%%%%%

\section{Dissipative evolution in Gaussian environments}\label{master}

We will consider the dissipative evolution of the infinite dimensional 
$n$-mode bosonic system coupled to an environment modeled by a continuum of 
oscillators. The couplings and the baths interacting with different modes
will be uncorrelated and generally different,
each bath being made up by a different continuum of oscillators..
The bath associated to mode $i$
will be labeled by the subscript $i$.
The dynamics of the system and of the reservoirs
is described by the following interaction Hamiltonian
\be
H_{int}=\sum_{i=1}^{n}
\int [w_i(\omega)a_i^{\dag}b_i(\omega)+w_i(\omega)^{*}a_ib_i^{\dag}(\omega)]
\,{\rm d}\,\omega \, , \label{coupling}
\ee
where $b_i(\omega)$ stands for the annihilation operator of the
$i$th bath mode
labeled by the variable $\omega$, whereas $w_i(\omega)$
represents the coupling of such a mode to the mode $i$ of the system
(taking into account the density of environmental modes).
The state of the bath is assumed to be
stationary. Under the Markovian approximation, such a coupling
gives rise to a time evolution ruled by the following
master equation (in interaction picture) \cite{walls}
\be
\dot\varrho\;  = \; \sum_{i=1}^{n}
\frac{\gamma_i}{2}\Big(N_i \: L[a_i^{\dag}]\varrho
+(N_i+1)\:L[a_i]\varrho - %\nonumber \\
  M_i^{*}\:D[a_i]\varrho + M_i
D[a_i^{\dag}]\varrho \Big)
\label{rhoev} \, ,
\ee
where the dot stands
for time--derivative,
the Lindblad superoperators are defined as
$L[\hat{o}]\varrho \equiv  2 \hat{o}\varrho \hat{o}^{\dag} -
\hat{o}^{\dag} \hat{o}\varrho -\varrho \hat{o}^{\dag} \hat{o}$ and
$D[\hat{o}]\varrho \equiv  2 \hat{o}\varrho \hat{o}
-\hat{o} \hat{o}\varrho -\varrho \hat{o} \hat{o}$,
the couplings are $\gamma_i=2\pi w_i^{2}(0)$,
whereas the coefficients $N_i$ and $M_i$
are defined in terms of the correlation functions
$\langle b_i^{\dag}(0)b_i(\omega) \rangle = N_i\delta(\omega)$ and
$\langle b_i(0)b_i(\omega) \rangle = M_i\delta(\omega)$,
where averages are computed over the state of the bath.
The requirement of positivity of the density matrix at any given time imposes
the constraint $|M_i|^{2} \le N_i(N_i+1)$.
At thermal equilibrium, {\it i.e.}~for $M_i=0$, $N_i$
coincides with the average number of thermal photons
in the bath.
If $M_i\neq 0$ then the bath $i$ is said to be `squeezed', or phase-sensitive,
entailing reduced fluctuations in one field quadrature.
A squeezed reservoir
may be modeled as the interaction with a bath of
oscillators excited in squeezed thermal states \cite{sqbath1};
several effective realization of such reservoirs have been
proposed in recent years \cite{tomvit,cir}.
In particular,
in Ref.~\cite{tomvit} the authors show that a
squeezed environment can be obtained, for
a mode of the radiation field, by means of
feedback schemes relying on QND
`intracavity' measurements, capable of affecting
the master equation of the system \cite{wise}. More specifically,
an effective squeezed reservoir is shown to be
the result of a continuous homodyne
monitoring of a field quadrature,
with the addition of a feedback driving term, coupling the 
homodyne output current with another field quadrature
of the mode.\par  
In general, the real parameters $N_i$ and the complex
parameters $M_i$ allow for the description 
of the most general single--mode Gaussian reservoir, 
fully characterized by its covariance 
matrix $\gr{\sigma}_{i\infty}$, given by 
\be
\gr{\sigma}_{i\infty}=\left(\begin{array}{cc}
\frac12+N_i+\,{\rm Re}\,M_i & {\rm Im}\,M_i \\
{\rm Im}\,M_i & \frac12+N_i+\,{\rm Re}\,M_i
\end{array}\right) \; . \label{envi}
\ee 
The non unitary evolution 
of the single mode system interacting 
with the reservoir $i$ can be seen as a quantum channel acting
on the original state.
The Gaussian state with null first moments and second moments
given by \eq{envi} constitutes the asymptotic state 
of such a channel
irrespective of the initial condition and, together with the coupling $\gamma_i$,
completely characterizes the channel.
Now, because of Williamson theorem any centered single mode Gaussian state
$\varrho$ referring to mode $i$ can be written as
\be
\varrho=S_{r_i,\varphi_i}^{\dag}\varrho_{\nu_i}S_{r_i,\varphi_i}\; , \label{sm}
\ee
where $S_{r_i,\varphi_i}$ will denote, from now on, the single mode squeezing operator
$S_{ii,r_i,\varphi_i}$.
This fact promptly provides
a more suitable parametrization of the asymptotic (or `environmental')
state (which is indeed a centered single-mode Gaussian state), given
by the following equations \cite{paris03}
\begin{eqnarray}
\mu_{i\infty}&=&\frac{1}{\sqrt{(2N_i+1)^{2}-4|M_i|^{2}}}
\: , \label{purasi} \\
&& \nonumber \\
\cosh(2r_i)&=&\sqrt{1+4\mu_{i\infty}^{2}|M_i|^{2}}
\: , \label{squizasi} \\
&& \nonumber \\
\tan(2\varphi_i)&=&-\tan\left({\rm Arg}{M_i}\right)
\: . \label{phiasi}
\end{eqnarray}
The quantities $\mu_{i\infty}$, $r_i$ and 
$\varphi_i$ are, respectively, the purity, 
the squeezing parameter and the squeezing angle of 
the squeezed thermal state of 
the bath. The quantity $\mu_{i\infty}$ is determined, in terms of the parameters 
of \eq{sm}, by $\mu_{i\infty}=1/(2\nu_i)$: the purity of a Gaussian state is
fully determined by the broadness of the thermal state providing its normal
mode decomposition.\par
Eq.~(\ref{rhoev}) is equivalent to the following
diffusion equation for the characteristic function
$\chi$ in terms of the quadrature variables $x_i$ and $p_i$
of mode $i$ \cite{barnett}
\be
\dot{\chi}(X,t)  =  -\sum_{i=1}^{n}\frac{\gamma_i}{2}\Bigg[
(x_i \; p_i){\partial_{x_i} \choose \partial_{p_i}}
+(x_i \; p_i) \gr{\sigma}_{i\infty}{x_i \choose p_i}
\Bigg] \chi(X,t)\, .
\label{diff}
\ee
It is easy to verify that, for any initial condition
$\chi_{0}(X)$, the following expression solves Eq.~(\ref{diff})
\be
\chi(X,t)=\chi_0 (\Gamma(t) X)
\,{\rm e}^{-\frac12 X^T \gr{\sigma}_{\infty}(t) X} \; . \label{solu}
\ee
with the $2n\times 2n$ real matrices
$\Gamma$ and $\gr{\sigma}_{\infty}(t)$ defined as
$$\Gamma(t)=\bigotimes_i \,{\rm e}^{-\frac{\gamma_i}{2}t}
{\mathbbm 1}_2\, ,$$
$$\gr{\sigma}_{\infty}(t)=\oplus_i \gr{\sigma}_{i\infty}(1-\,{\rm e}^{-\gamma_i t}) \, .$$
We mention that \eq{rhoev} can be equivalently recast as a Fokker Planck equation for the
Wigner function \cite{barnett}, as follows
\be
\dot{W}(X,t)=\sum_{i=1}^{n}\frac{\gamma_i}{2}\Bigg[
(\partial_{x_i}\; \partial_{p_i}) {x_i\choose p_i} + (\partial_{x_i}\; \partial_{p_i})
\gr{\sigma}_{i\infty}{\partial_{x_i}\choose \partial_{p_i}}
\Bigg] W(X,t) \, .
\label{fokpla}
\ee

Let us now consider a $n$-mode
Gaussian state with CM $\gr{\sigma}_0$ and first moments $X_0$
as initial condition in the Gaussian noisy channel. Inserting \eq{gauchar} in \eq{solu}
shows that the evolving state maintains its Gaussian character and is therefore
characterized by the action of dissipation on the first and second moments.
At time $t$ one has
\bea
X(t)&=&\Gamma (t) X_0 \; , \label{1evo}\\
\gr{\sigma}(t)&=&\Gamma(t)\gr{\sigma}_0 \Gamma(t)+\gr{\sigma}_{\infty}(t) \, .
\label{2evo}
\eea
In particular, focusing on second moments, \eq{2evo} is, at any given time $t$, a relevant
example of Gaussian completely positive map. Actually,
in a more general framework, it can be shown that any
evolution resulting from the reduction of a symplectic evolution on a larger Hilbert space
can be described, in terms of second moments, by
\be
\gr{\sigma}\rightarrow X^T \gr{\sigma} X + Y \, , \label{gencha}
\ee
where $X$ and $Y$ are $2n\times 2n$ real matrices fulfilling
$Y+i\Omega-iX^T\Omega X\ge 0$ \cite{book, canali}.
Viceversa, any evolution of this kind may be interpreted
as the reduction of a larger symplectic evolution.

As a last remark about the dissipative evolution under the master equation (\ref{rhoev}),
we point out an interesting general feature concerning a single-mode
non-squeezed bath, characterized by its asymptotic purity $\mu_{\infty}$.
Let us consider the evolution in such a channel of an intial pure non Gaussian state
(whose Wigner function necessarily takes negative values).
It can be shown by a beautiful geometric argument \cite{brodier} that
the instant $t_{nc}$ at which the state's Wigner
function gets non negative, so that the nonclassicality of the state quantified by its
negative part $\xi$ becomes null, does not depend on the chosen state at all.
Such a time (that is also referred to as `positive time') reads
\be
t_{nc}=\frac{1}{\gamma}\ln(1+\mu_{\infty}) \; .
\label{tnc}
\ee
In section \ref{numero} we will provide a simple proof of this result for an intial number state.
%%%%%%%%%%%%%%%%%%%%%%%%%%%%%%%%%%%%%%%%%%%%%%%%%%

\section{Single-mode Gaussian states}\label{1mode}

The set of single-mode Gaussian states can be regarded as the simplest 
continuous variable arena
in which the decay of quantum coherence can be examined.
The evolution of single-mode Gaussian states in thermal reservoirs has been extensively
addressed in Ref.~\cite{marian}, while their dissipative evolution 
under a general Lindblad equation has been studied in Ref.~\cite{rajapla}.
Ref.~\cite{paris03} contains many of the results
which will be here reviewed for phase-sensitive baths.
Both the purity and the nonclassical depth of Gaussian states are completely determined
by their CM $\gr{\sigma}$, on which we will thus focus. Exploiting
again \eq{sm}, we parametrize the $2\times 2$ CM $\gr{\sigma}$
through the parameters $\mu$, $r$ and $\varphi$, according to
\bea
\sigma_{11}&=&\frac{1}{2\mu}[\cosh(2r)-\sinh(2r)\cos(2\varphi)]\, ,\nonumber \\
\sigma_{22}&=&\frac{1}{2\mu}[\cosh(2r)+\sinh(2r)\cos(2\varphi)]\, ,\label{1para}\\
\sigma_{12}&=&\frac{1}{2\mu}\sinh(2r)\sin(2\varphi)\nonumber\, .
\eea
Notice that the purity $\mu$ characterizes the CM according to \eq{purgau}.
The evolution in a channel characterized by $\gamma$, $\mu_\infty$, $r_\infty$ and
$\varphi_\infty$ of an initial state parametrized by $\mu_0$, $r_0$ and
$\varphi_0$ is provided by the single mode ($n=1$) instance of \eq{2evo}.
Such an equation, together with the parametrization of Eqs.~(\ref{1para}) can be
exploited to promptly achieve the time evolution
of the parameters $\mu$, $r$ and $\varphi$, yielding
\be
\begin{split}
\mu(t)=&\mu_{0}\bigg[\frac{\mu_{0}^{2}}{\mu_{\infty}^{2}}\left(1-
{\rm e}^{-\gamma t}\right)^{2} \, + \,
{\rm e}^{-2\gamma t}
+2\frac{\mu_{0}}{\mu_{\infty}}\Big(\cosh(2r_{\infty})\cosh(2r_{0})\label{1muevo}\\
&+\sinh(2r_{\infty})\sinh(2r_{0})\big(\cos(2\varphi_{\infty}-2\varphi_{0})
\big)
\Big)
\left(1-{\rm e}^{-\gamma t}\right){\rm e}^{-\gamma t}
\bigg]^{-\frac12} \, ,
\end{split}
\ee
\begin{equation}
\frac{\cosh[2r(t)]}{\mu(t)}=
\frac{\cosh(2r_{0})}{\mu_{0}}\,{\rm e}^{-\gamma t}+
\frac{\cosh(2r_\infty)}{\mu_{\infty}}\left(1-
{\rm e}^{-\gamma t}\right)
{\rm
\, ,} \label{1revo}
\end{equation}
\begin{equation}
\tan[2\varphi(t)]=\frac{\sinh(2r_{0})\sin(2\varphi_{0}){\rm e}^{-\gamma t}
-\sin(2\varphi_\infty)\frac{\mu_{0}}{\mu_\infty}
\left(1-{\rm e}^{-\gamma t}\right)}
{\sinh(2r_{0})\cos(2\varphi_{0}){\rm e}^{-\gamma t}
-\cos(2\varphi_\infty)\frac{\mu_{0}}{\mu_\infty}
\left(1-{\rm e}^{-\gamma t}\right)}{\rm
\: .} \label{1phievo}
\end{equation}
First of all, according to intuition, the purity $\mu(t)$ is an
increasing function of the input purity $\mu_0$: this complies with
a general fact about output purities of channels, which are maximized by pure states,
due to their convexity \cite{canali}.
Moreover, it is immediate to see from \eq{1muevo} that in a non squeezed thermal
bath ({\em i.e.~}for $r_\infty=0$), the purity is maximum at any given time $t$
for $r_0=0$: the output purity of such a channel is maximized for
$r_0=0$, that is for a coherent input
state.\footnote{This is a particular instance of a more general result concerning the output
purity of Gaussian bosonic channels of the form of \eq{gencha} \cite{canali, lloyd}.}
In the theory of measurement,
the fact that coherent states yield the minimal entropic production
-- under non unitary evolution in thermal reservoirs -- is well known and
selects such states as privileged `pointer states' in measurement processes
\cite{zurek93,venugopalan}.\footnote{Notice that the couplings to the bath
of oscillators typically considered in these cases are
not symmetric under the exchange of the two quadratures: this is the reason why,
at very small times, some squeezing provides greater purity in such models \cite{zurek93}.
On the other hand, the coupling we consider in \eq{coupling} is manifestely
symmetric in $\hat{x}_i$ and $\hat{p}_i$.}

For phase-sensitive bath, with $r_\infty,\,\varphi_\infty\neq 0$, the
purity $\mu(t)$ is maximized for $r_0=r_\infty$ and $\varphi_{0}=\varphi_\infty+\pi/2$.
This should be expected: in fact, in terms of the single mode
squeezing operator $S_{r,\varphi}$ entering
\eq{sm}, this means that the optimal input state is countersqueezed with respect to the bath,
since $S_{r,\varphi+\frac\pi2}=S^{-1}_{r,\varphi}$. Indeed, since the purity 
is invariant under unitary transformation, such a result is just a consequence 
of the fact that the evolution in non squeezed baths is optimized by coherent inputs.\footnote{More
formally, one can exploit the invariance of the purity under $Sp_{(2,\R)}$ and
bring the CM $\gr{\sigma}_\infty$ of the bath in Williamson standard form: in these
canonical basis of phase space the channel is non squeezed and coherent states (with CM
$\gr{\sigma}_0={\mathbbm 1}_2/2$) maximize the purity. To go back to the original
canonical basis one has to apply the inverse symplectic transformation: this explains
the previous result about optimization.}
%%%%%%%%%%%%%%%%%%%%%%%%%%%%%%%%%%%%%%%%%%%%%%%%%%
\begin{figure}[tb!]
\begin{center}
\includegraphics[scale=1]{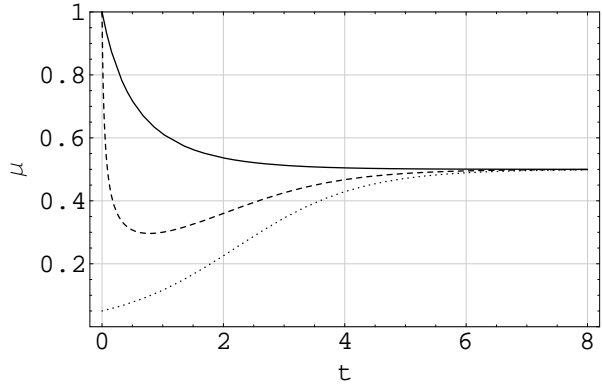}
\caption{\sf \footnotesize Evolution of the purity of various Gaussian states in a channel 
with $\mu_{\infty}=0.5$ and $r_{\infty}=0$. The continuous line refers to an 
initial pure coherent state ($\mu_0=1$, $r_0=0$), the dashed line refers to 
a squeezed vacuum ($\mu_0=1$, $r_0=1.5$) while the dotted line refers to 
a thermal state with $\mu_0=0.05$ and $r_0=0$.\label{pur1m}}
\end{center}
\end{figure}
%%%%%%%%%%%%%%%%%%%%%%%%%%%%%%%%%%%%%%%%%%%%%%%%%%
The optimal evolution of purity, plotted in Fig.~\ref{pur1m}, is simply obtained by inserting
$r_0=r_{\infty}=0$ in \eq{1muevo}.\par
For an initial squeezed input with squeezing parameter $r_0$ 
in a thermal bath with $r_{\infty}=0$ (or, more generally, for an
initial state with relative squeezing $r_0-r_{\infty}\neq 0$) the purity
$\mu(t)$ may display a local minimum. The condition for the appearance of such
a minimum can be simply derived by differentiating \eq{1muevo} and turns out
to be $r_0>\max\left[\mu_0/\mu_{\infty},\mu_{\infty}/\mu_0\right]$; the time
$t_{min}$ at which the minimum is attained can be exactly determined as
\be
t_{min}=-\frac{1}{\gamma}\ln\left[\frac{\frac{\mu_0}{\mu_{\infty}}-\cosh(2r_0)}{\frac{\mu_0}{\mu_{\infty}}+
\frac{\mu_{\infty}}{\mu_0}-2\cosh(2r_0)}\right] \; .
\ee
The time $t_{min}$ provides a good characterization of the decoherence time 
of the squeezed state: during the initial steep fall of the purity the
coherence and the information contained in the initial state are irreversibly
spread in the environmental modes. The subsequent revival of the purity is just
a result of the driving of the state of the system 
towards the (asymptotically reached) environmental one.

\begin{figure}[tb!]
\begin{center}
\includegraphics[scale=1]{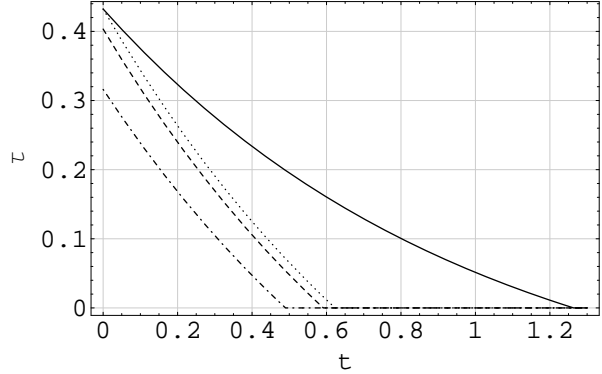}
\caption{\sf \footnotesize Evolution of the nonclassicality $\tau$ in various channels with 
$\mu_{\infty}=0.5$. The dotted and the continuous lines refer to an initial 
squeezed vacuum ($\mu_0=1$, $r_0=1$) evolving, respectively, in a non squeezed
channel (dotted line) and in a channel with $r_{\infty}=0.2$ (continuous line). 
The dotted line refers to an initial state with $\mu_0=0.7$ and $r_0=1$ and the
dot-dashed line to an initial state with $\mu_0=1$ and $r_0=0.5$.
\label{nc1m}}
\end{center}
\end{figure}

Concerning the nonclassical depth, the smallest eigenvalue $u$ of a single mode Gaussian state
is simply found in terms of $\mu$, $r$ and $\varphi$ as $u=\,{\rm e}^{-2r}/(2\mu)$. 
Inserting such a result into \eq{ncgau} gives the following equation for the non classical 
depth $\tau$ of a single mode Gaussian state
\be
\tau=\max\left[\frac{1-\frac{{\rm e}^{-2r}}{\mu}}{2},0\right] \; . \label{1ncgau}
\ee
Let us define the quantity $\kappa(t)$ as
\[
\kappa(t)=\frac{\cosh(2r_{0})}{\mu_{0}}\,{\rm e}^{-\gamma t}+
\frac{\cosh(2r_\infty)}{\mu_{\infty}}\left(1-
{\rm e}^{-\gamma t}\right) \, .
\]
Notice that $\kappa$ is an increasing function of $r_0$ and a decreasing function of $\mu_{0}$.
After some algebra, Eqs.~(\ref{1revo}) and (\ref{1ncgau}) yield the following result 
for the exact time evolution of the nonclassicality of a single mode Gaussian state
\be
\tau(t)=\frac{1-\kappa(t)+\sqrt{\kappa(t)^{2}-\frac{1}{\mu(t)^2}}}{2} \, .
\ee 
Such a function increases with both $\mu(t)$ and $\kappa(t)$. 
The choice of the input phase of the squeezing which maximizes $\tau(t)$ at any time 
is again $\varphi_0=\varphi_{\infty}+\pi/2$, maximizing
the purity. 
The maximization of $\tau(t)$ in terms of the other parameters of the initial state is
the result of the competition of two different effects. Let us consider 
$r_0$: on the one hand a
squeezing parameter $r_0$ matching the squeezing $r_{\infty}$ maximizes the purity
thus delaying the decrease of $\tau(t)$; on the other hand, a bigger value of
$r_0$ obviously yields a greater initial $\tau(0)$.  
However the numerical analysis, summarized in Fig.~\ref{nc1m}, 
unambiguously shows that, in non
squeezed baths, the nonclassical depth increases with increasing squeezing
$r_0$ and purity $\mu_0$, as one should expect.
%%%%%%%%%%%%%%%%%%%%%%%%%%%%%%%%%%%%%%%%%%%%%%%%%%
%%%%%%%%%%%%%%%%%%%%%%%%%%%%%%%%%%%%%%%%%%%%%%%%%%

\section{Schr\"odinger cats}\label{cats}

We consider now the following coherent normalized superposition of single mode
displaced squeezed states
\begin{equation}
|\beta_{0},\theta\rangle \equiv
\frac{|\beta_{0}\rangle + \;{\rm e}^{i\theta}|-\beta_{0}\rangle}
{\sqrt{2+2\cos(\theta)\;{\rm e}^{-2\|X_0\|^{2}}}}\: ,
\end{equation}
where $\ket{\beta_0}=S_{r_0,0}D_{X_0}\ket{0}$,
and address its evolution under the master equation (\ref{rhoev}).
The choice of a null phase in the operator $S_{r_0,0}$ is just a 
reference choice for phase space rotations.\par
This state is a relevant instance of cat-like state, {\em i.e.~}of 
coherent superposition of pure quantum states, whose macroscopic 
extension has been invoked by Schr\"odinger to illustrate some of the counterintuitive
features of quantum mechanics \cite{schroedinger}.
More recently,
the seminal proposal by Yurke and Stoler \cite{yusto}, besides spurring a great amount
of theoretical work aimed at optimizing the generation of cat-like states \cite{catgene}, lead
to the experimental realization
of mesoscopic ($\|X_0\|\simeq10$) superposition of Gaussian states of
the radiation field in cavity QED \cite{haroche96}.
The realization of superpositions of Gaussian motional states
of trapped particles
has been demonstrated as well \cite{monroecat}, together with the
experimental investigation of their
rates of decoherence \cite{myatt}.
On the theoretical side, many efforts have been done to understand and, possibly,
suggest methods to control the decoherence of such superpositions
\cite{walls85, kennedy, vitali, el-ora, gatti}.
Furthermore, we mention that
an accurate analysis, under the `quantum jump' approach,
of the decoherence of nonclassical quantum optical states
(encompassing both cat-like and number states)
can be found in Ref.~\cite{garraway}, where
it is also shown how nonclassical states may be the result of
proper dissipative evolutions.
Most of the results here reviewed can be found in
Ref.~\cite{gatti}.
\par
Let us define the matrices $\gr{R}=\,{\rm diag}\,(\,{\rm e}^{r_0},\,{\rm e}^{-r_0})$
(corresponding to the action of $S_{r_0,0}$ on the $2$-dimensional phase space), and 
$\gr{\sigma}_{0}=1/2 \gr{R}^2$. The Wigner function associated to 
the state $\ket{\beta_0}$ reads
\begin{eqnarray}
W_{\beta _{0},\theta}(X) &=&\frac{1}{4\pi (1+\cos(\theta)\,\mathrm{e}%
^{-\|X_0\|^2})\sqrt{{\rm Det}\,\boldsymbol{\sigma}_{0}}}
\nonumber\\
&&\times \left[ 
\,\mathrm{e}^{-\frac12(X^{T}-X_0^T{\bf R})
\boldsymbol{\sigma}_{0}^{-1}(X-%
{\bf R}X_0)}\right.  
+\left.\,\mathrm{e}^{-\frac12(X^{T}+X_0^T{\bf R})\boldsymbol{\sigma}_{0}%
^{-1}(X+{\bf R}X_0)}\right.\nonumber\\
&&+ \left.\,\mathrm{e}^{-\|X_0\|^2}\left( 
\,\mathrm{e}^{-\frac12(X^{T}-iX_0^T\omega{\bf R})\boldsymbol{\sigma}_{0}^{-1}
(X+i{\bf R}\omega X_0)+i\theta}+c.c.\right) \right] \; ,
\end{eqnarray}
consisting in the two Gaussian peaks at the phase space points $X_0$ and $-X_0$,
linked in phase space by the oscillating 
interference terms. Obviously, this Wigner function is non positive.
However, formally, such a function is just the sum of four 
displaced Gaussian terms.
The linearity of the considered dissipative evolution permits to simply 
solve the evolution of the cat state, by following the evolution of its four
Gaussian terms according to Eqs.~(\ref{1evo}, \ref{2evo}). One gets
\begin{equation}
\begin{split}
W_{\beta _{0},\theta}(X) =&\frac{1}{4\pi (1+\cos(\theta)\,\mathrm{e}%
^{-\|X_0\|^2})\sqrt{{\rm Det}\,\boldsymbol{\sigma}(t)}}\\
%\nonumber\\
&\times \left[ 
\,\mathrm{e}^{-\frac12(X^{T}-\,{\rm e}^{-\frac{\gamma}{2}t}X_0^T{\bf R})
\boldsymbol{\sigma}(t)^{-1}(X-\,{\rm e}^{-\frac{\gamma}{2}t}
{\bf R}X_0)}\right.\\  %\nonumber \\
&+\left.\,\mathrm{e}^{-\frac12(X^{T}+\,{\rm e}^{-\frac{\gamma}{2}t}
X_0^T{\bf R})\boldsymbol{\sigma}(t)
^{-1}(X+\,{\rm e}^{-\frac{\gamma}{2}t}{\bf R}X_0)}\right. \\%\nonumber\\
&+ \left.\,\mathrm{e}^{-\|X_0\|^2}\left( 
\,\mathrm{e}^{-\frac12(X^{T}-i\,{\rm e}^{-\frac{\gamma}{2}t}X_0^T\omega{\bf R})\boldsymbol{\sigma}(t)^{-1}
(X+i\,{\rm e}^{-\frac{\gamma}{2}t}
{\bf R}\omega X_0)+i\theta}+c.c.\right) \right] \; ,
\end{split}
\end{equation}
where $\gr{\sigma}(t)$ is given by \eq{1evo} with $\gr{\sigma}_0$ defined above.

%%%%%%%%%%%%%%%%%%%%%%%%%%%%%%%%%%%%%%%%%%%%%%%%%%
\begin{figure*}[tb!]
\begin{center}
%\hspace*{-20mm}
 \subfigure[\label{wigcat1}]
{\includegraphics[width=6cm]{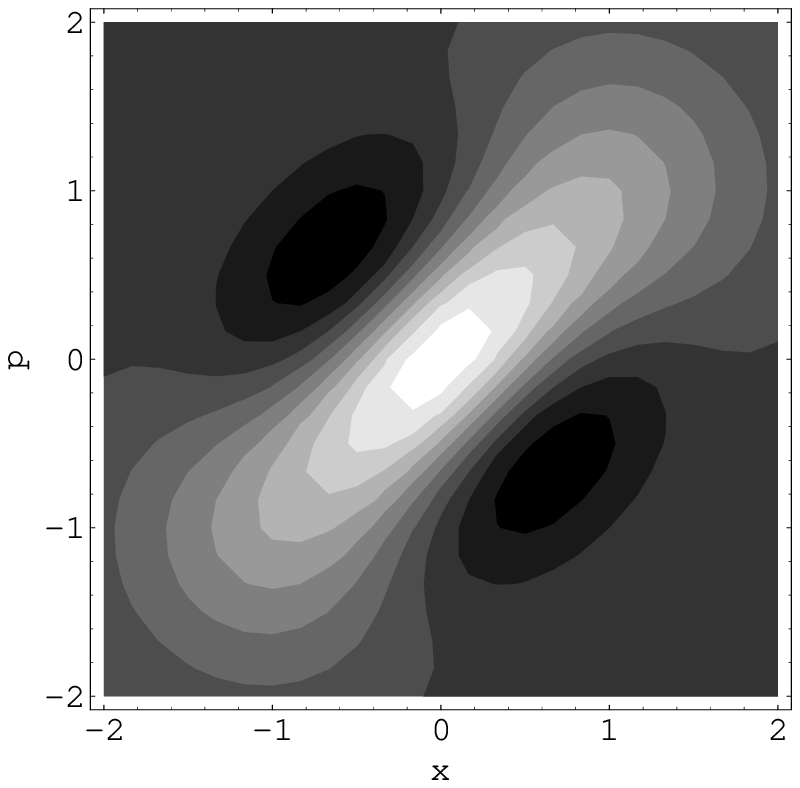}}
%\hspace{5mm}
\subfigure[\label{wigcat2}]
{\includegraphics[width=6cm]{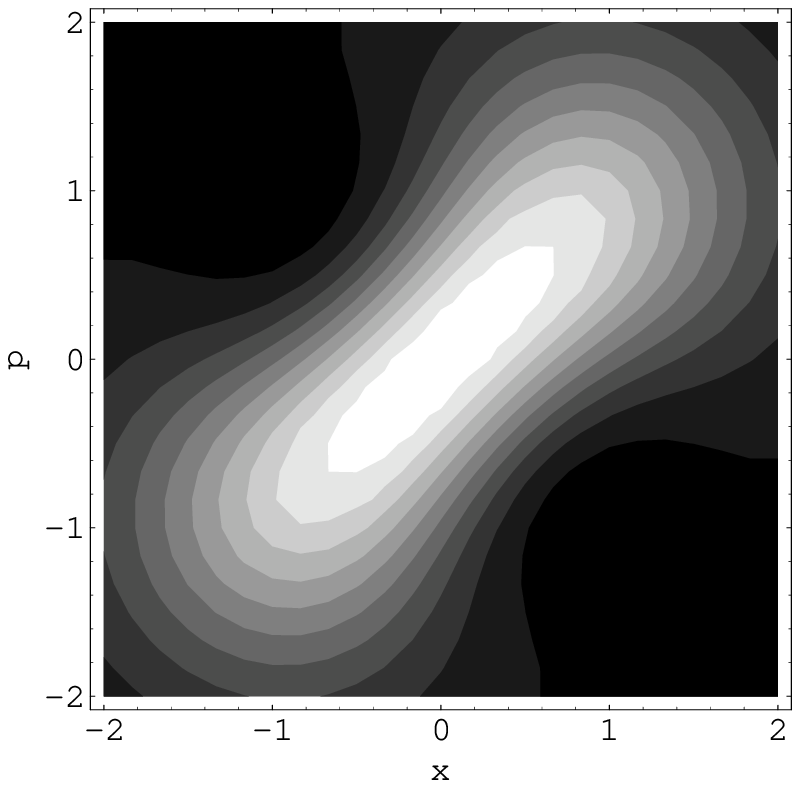}}
\subfigure[\label{wigcat3}]
{\includegraphics[width=6cm]{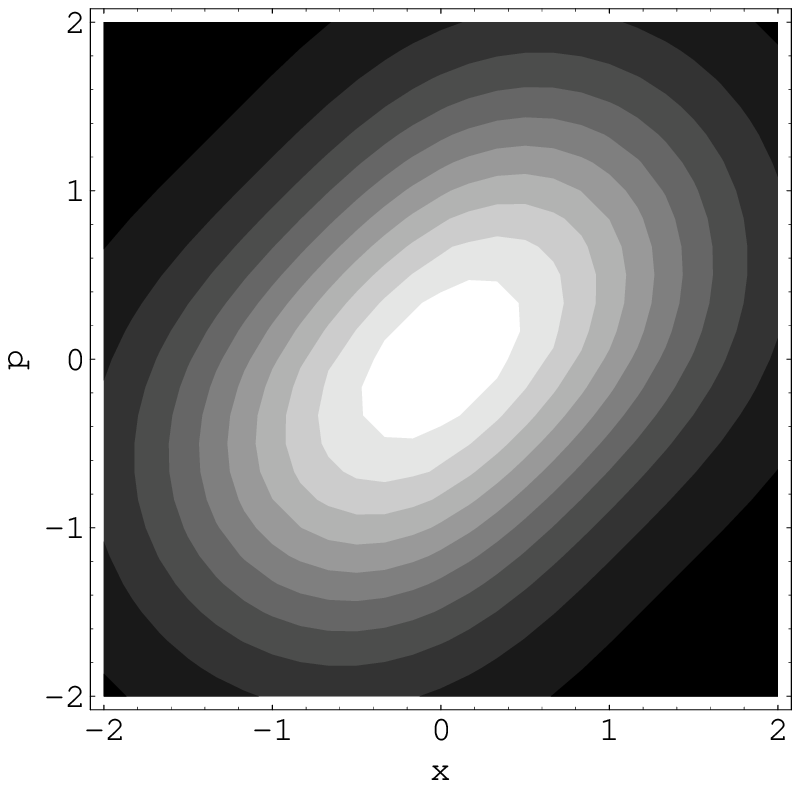}}
% \hspace{7mm}
\subfigure[\label{wigcat4}]
{\includegraphics[width=6cm]{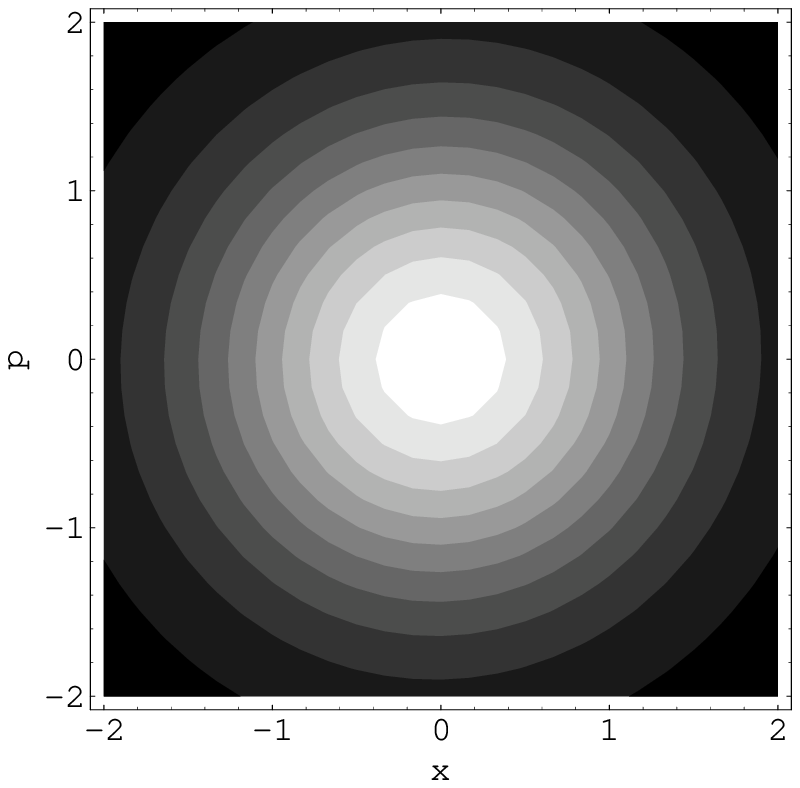}}
\end{center}
\caption{\sf \footnotesize Evolution in phase space of the Wigner function of an initial
non squeezed cat-like state with $X_0^T=(1\; 1)$ and $\theta=0$ in a thermal channel with 
$\mu_{\infty}=0.5$ at times $t=0$ (a), $t=\gamma^{-1}/2$ (b), $t=\gamma^{-1}$ (c)
and $t=4\gamma^{-1}$ (d). Darker colors stand for lower values, the scale of each plot 
is normalized. The negative lobes (in which the Wigner function takes negative values), 
evident in subfigure (a) are already disappeared in subfigure (b). Actually, the positive 
time $t_{nc}$ of such a reservoir is $t_{nc}\simeq 0.4 \gamma^{-1}$ (see \eq{tnc}). 
\label{wigcat}}
\end{figure*}
%%%%%%%%%%%%%%%%%%%%%%%%%%%%%%%%%%%%%%%%%%%%%%%%%%

Figure \ref{wigcat} provides a relevant example of dissipation of a cat 
state in a thermal environment, isotrope in phase space. The negative part $\xi$  
of the Wigner function reaches the value $0$ at a time $t_{nc}\simeq 0.4\gamma^{-1}$, 
in agreement with \eq{tnc}. As already mentioned, this time is feature of the 
bath and does not depend on the initial pure (non Gaussian) state.

\begin{figure}[tb!]
\centering
\includegraphics[scale=1]{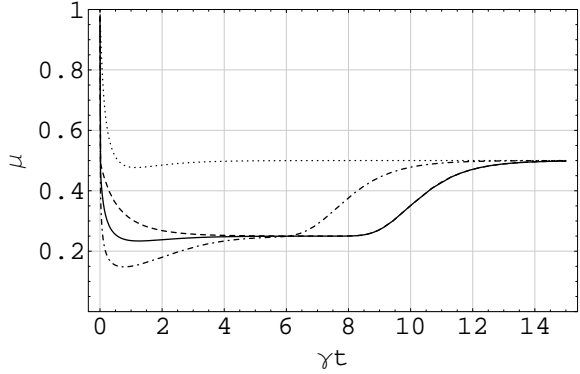}
\caption{\sf \footnotesize Evolution of the purity of initial cat-like states.
The asymptotic purity of the channel is $\mu_{\infty}=0.5$. 
The dotted line refers to a cat state with $X_0^T=(1,\; 1)$ in a non squeezed
channel. The dashed and the continuous lines refer to an initial non squeezed cat with 
$X_0^T=(100, \; 100)$ evolving in a non squeezed channel and in a channel with 
$r_{\infty}\simeq 0.88$ and $\varphi_{\infty}=-\pi/8$. 
The dot-dashed line refers to an initial state with $X_0^T=(10,\; 10)$ and 
$r_0=2$ evolving in a non squeezed channel.\label{purcat}}
\end{figure}

\begin{figure}[tb!]
\centering
\includegraphics[scale=1]{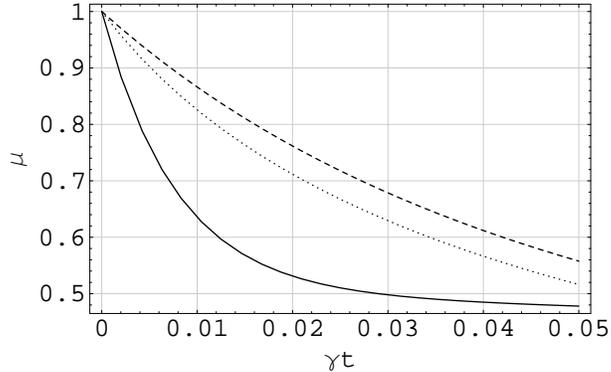}
\caption{\sf \footnotesize Comparison between the evolution at short times ({\em i.e.~}for $\gamma t_{dec\simeq 1}$)
of the purity of an initial non squeezed cat
(continuous line) and that of squeezed cats with optimal choice of the optical phase. 
In all instances $X_0^T=(4, \; 4)$, $\mu_{\infty}=0.5$ and $\theta=0$. 
The dashed line refers to a cat state with $r_0=1$, whereas the dotted line refers to a 
state with $r_0=1.5$. The decoherence time of such cats can be estimated as 
$t_{dec}\simeq0.03\gamma^{-1}$, in good agreement with the decrease of 
purtiy of the non squeezed cat. The remarkable delay of decoherence induced by squeezing the
cat can be appreciated, especially at $t\simeq t_{dec}$.\label{purcats}}
\end{figure}

The exact analytical expression of the 
purity of the evolving superposition is easily determined 
by Gaussian integrations,
according 
to \eq{wigpur}

\begin{eqnarray}
\mu_{\beta_{0},\theta}(t) &= &\left(8(1+\cos(\theta)\,
{\rm e}^{-\|X_0\|^2})^{2}
\sqrt{{\rm Det}\,\boldsymbol{\sigma}(t)}\right)^{-1}\nonumber\\
&&{\times \Bigg[2}\left(1+{\rm e}^{-\,{\rm e}^{-\gamma t}X_{0}^{T}
{\bf S}(t)X_{0}}\right)+
2\,{\rm e}^{-2\|X_0\|^2}\left(\cos(2\theta)+{\rm e}^{\,{\rm e}^{-\gamma t}
X_{0}^{T}
{\bf T}(t)X_{0}}\right)\nonumber\\
&&{+ 
4\,{\rm e}}^{-\|X_0\|^2}\cos(\theta)
\left( {\rm e}^{-\,{\rm e}^{-\gamma t}X_0^T \gr{J} X_0 
\,{\rm Tr}[\gr{J}\gr{S}(t)]^{*}/4} + c. c.\right)\Bigg] \, ,
\label{catmuevo}
\end{eqnarray}
with
\begin{equation}
{\bf S}(t) \equiv {\bf R}\boldsymbol{\sigma}(t)^{-1}{\bf R} \, , \quad
{\bf T}(t)\equiv ({\rm Det}\,\gr{\sigma})^{-1}{\bf S}(t)^{-1}\, ,\quad
{\bf J}\equiv \left(\begin{array}{cc}
1&i\\
i&-1\end{array}\right) \, .
\end{equation}
\eq{catmuevo} shows that the decoherence rate increases with the `dimension'
of the cat, quantified by $\|X_0\|$; in the limiting instance $X_0=0$, \eq{catmuevo}
reduces to \eq{1muevo} for an initial squeezed vacuum, which decoheres more slowly
than the equally squeezed cat-like states. Moreover, in general, the terms depending 
on the coherent phase $\theta$ are suppressed by exponential terms of the form 
$\exp(-\|X_0\|^2)$, so that the decoherence rate in terms of the purity is only
slightly influenced by the choice of $\theta$. 
Examples of decoherence of cat states 
can be seen in Fig.~\ref{purcat}. 
In all the instances the purity displays a fast initial fall,
during which all the coherence and the information of the pure cat-like state are lost.
The typical time scale in which the minimum of the purity is attained is in good agreement
with the estimate $t_{dec}=\gamma^{-1}/2\|X_0\|^2$, holding for the decoherence
time of a cat state in a thermal bath \cite{myatt}.
As can be shown analytically \cite{gatti}, the phase space direction of the cat, determined by the
angle $\xi_0=\arctan(x_0/p_0)$, providing the maximal delay of decoherence
at short times ({\em i.e.~}for $\gamma t\simeq1$) is given by
$\xi_0=\varphi_0+\pi/2$ for a squeezed cat in a 
non squeezed bath or, equivalently, by $\xi_0=\varphi_{\infty}$ for a non squeezed cat
in a squeezed bath. These two instances are, as already noted, unitarily
equivalent. In general, the evolution of the purity of an initial state
in a squeezed reservoir is identical to the one of the counter-squeezed initial state in a 
thermal reservoir. Therefore, the same protection against decoherence granted by squeezing the 
bath can be achieved by orthogonally squeezing the initial state. Indeed, with the optimal,
previously discussed, locking of the optical phase, an optimal value of the squeezing $r_0$ 
maximizing the purity in non squeezed baths does exist. As illustrated by Fig.~\ref{purcats},
squeezing the initial cat (or the bath) can provide a significant delay of the complete
decoherence of the cat state, better preserving the interference fringes in phase space.  

%%%%%%%%%%%%%%%%%%%%%%%%%%%%%%%%%%%%%%%%%%%%%%%%%%
%%%%%%%%%%%%%%%%%%%%%%%%%%%%%%%%%%%%%%%%%%%%%%%%%%

\section{Number states}\label{numero}

As a last example of single mode state we quantify the decoherence of 
number states $\ket{n}\bra{n}$. 
Such states can be considered as probe of fundamental quantum mechanical features
and are also required in several quantum communications tasks \cite{crypto,secom}.
Different methods for the generation of Fock states have been proposed,
both for traveling-wave and cavity fields. For traveling-wave fields, 
these methods are principally based on tailored nonlinear interactions
\cite{non}, conditional measurements \cite{con}, state filtering 
\cite{fockfil} or state engineering 
\cite{eng}.  
A further possibility to generate
number states 
with high fidelities by atom-field interactions in
high-$Q$ cavities
has been recently suggested 
\cite{serra}. 
The actual experimental generation in quantum optical settings 
seems to be at hand, by both deterministic 
\cite{walther,brown2003} and 
probabilistic (`post-selective') schemes \cite{kurizki} (and the techniques to realize 
such states for motional degrees of freedom are well mastered \cite{wineland}),
even if the numerical analysis 
suggests that environmental decoherence could still hamper the very possibility of
generating pure number states \cite{nayak}. These reasons motivated an accurate 
investigation of the decoherence rate of number states, carried out in Ref.~\cite{num}.
We review such results, adding the analysis of the nonclassicality of the evolving states.

The characteristic function $\chi_n$ associated to the state $\ket{n}\bra{n}$ is promptly 
found and reads \cite{barnett}
\be
\chi_n(X)=\bra{n}D_\alpha\ket{n}=\,{\rm e}^{-\frac{\|X\|^2}{2}}
L_n (\|X\|^2) \, , \label{ini}
\ee
where $L_n$ is the Laguerre polynomial of order $n$:
$L_n (x)=\sum_{m=0}^{n}\frac{(-x)^m}{m!}{n \choose m}$. 
So that, exploiting \eq{solu}, one at once finds the evolution of such an initial state in the  
channel
\be
\chi_n(t)=L_n\left(\frac{\|X\|^2}{2}\,{\rm e}^{-\gamma t}\right)
\,{\rm e}^{-\frac12 X^T \gr{\sigma}(t) X} \, , \label{solun}
\ee
with
\be
\gr{\sigma}(t)=\frac{\mathbbm 1}{2}\,{\rm e}^{-\gamma t}+
\gr{\sigma}_{\infty}(1-\,{\rm e}^{-\gamma t}) \; .
\ee
According to \eq{wigpur} one can then determine the purity $\mu_{n}(t)$ of the
evolving number state \cite{tafeln}
\be
\mu_{n}(t)=\,{\rm e}^{\gamma t}
\int_{0}^{\infty}\,{\rm e}^{-\xi s}
L_n(s)
I_0\left(\frac{|\sinh(2r_{\infty})|}{2\mu_{\infty}}
(\,{\rm e}^{\gamma t}-1)s\right)\,{\rm d}s \, , \label{pursq}
\ee 
where $I_{0}(x)=J_{0}(ix)=
\sum_{k=0}^{\infty}\frac{x^{2k}}{(2^k k!)^2}$ is the  
zero order modified Bessel function of the first kind and 
\[
\xi=\frac{{\rm e}^{\gamma t}+\mu_{\infty}-1}{\mu_{\infty}} \; .
\]
For a thermal channel, with $r_{\infty}=0$, such an expression can be further
simplified to achieve an exact analytical expression for the purity, yielding \cite{tafeln}
\be
\mu_n(t)=\,{\rm e}^{\gamma t}\frac{(\xi-2)^2}{\xi^{n+1}}
P_n\left(1+\frac{2}{\xi^2-2\xi}\right) \, , \label{purter}
\ee
where $P_n$ is the Legendre polynomial of order $n$: $
P_n(x)=\frac{1}{2^n n!}\frac{{\rm d}^n}{{\rm d}\,x^n}(x^2-1)^n$.
Again, we point out that the squeezing of the bath has the same effect on the
purity as the counter squeezing of the initial number state, amounting
to consider a `squeezed number state'. The numerical analysis of \eq{pursq}
at short times (for $\gamma t\lesssim 1$) 
shows that $\mu_n(t)$ is a decreasing function of $r_{\infty}$: the 
squeezing of the bath does not help to preserve the coherence of number states. 
Also, the purity at any given time is a decreasing function of $n$: number states
of higher order are more fragile and decoheres faster.

Let us now deal with the evolution of the negative part $\xi$ of a number
state $\ket{n}$, quantifying the decoherence effect on the nonclassical 
features of the state. 
The initial value of such a quantity increases with increasing $n$ 
(higher order number states are regarded as `less classical' by this indicator).
Subsequently, during the dissipation in the bath, the negative part $\xi$ 
decreases up to a time $t_{nc}$ -- determined by \eq{tnc} -- 
at which it reaches the values $0$ and 
the nonclassical features of the state related to $\xi$ are erased. 
Interestingly, a direct determination of the time $t_{nc}$ can be easily 
provided for the
relevant instance of number states evolving in non squeezed thermal baths
(with $r_{\infty}=0$). In such a case, the spherically symmetric 
characteristic 
function of \eq{solun} for $r_{\infty}=0$ can be Fourier transformed
to get the Wigner function $W_n(t)$
\be
W_n(t)=\frac{\eta(t)^n}{\pi \zeta(t)^{n+1}}\,{\rm e}^{-\frac{\|X\|^2}
{\zeta(t)}}L_n\left[\frac{-2\,{\rm e}^{-\gamma t}\|X\|^2}
{\zeta(t)\eta(t)}\right] \, ,\label{wign}
\ee 
with
\[
\zeta(t)=\frac{1}{\mu_{\infty}}\left[1-(1-\mu_\infty)
\,{\rm e}^{-\gamma t}\right] \quad {\rm and}\quad
\eta(t)=\frac{1}{\mu_{\infty}}\left[1-(1+\mu_\infty)
\,{\rm e}^{-\gamma t}\right] \, .
\]
Since Laguerre polynomials of any order have positive roots and are always 
positive for negative arguments, \eq{wign} implies that
the time $t_{nc}$ is determined by the
condition $\eta(t_{nc})=0$, yielding 
$t_{nc}=\gamma^{-1}\ln(1+\mu_{\infty})$.
This result is just a specific instance of \eq{tnc}, which can be applied at any 
pure non Gaussian initial state.
It can also be found in Ref.~\cite{janszkim}, where the remarkable independence of the 
time $t_{nc}$ on the order $n$ of the number state had already been stressed.
%%%%%%%%%%%%%%%%%%%%%%%%%%%%%%%%%%%%%%%%%%%%%%%%%%
\begin{figure*}[tb!]
\begin{center}
%\hspace*{-20mm}
 \subfigure[\label{wignum1}]
{\includegraphics[width=6cm]{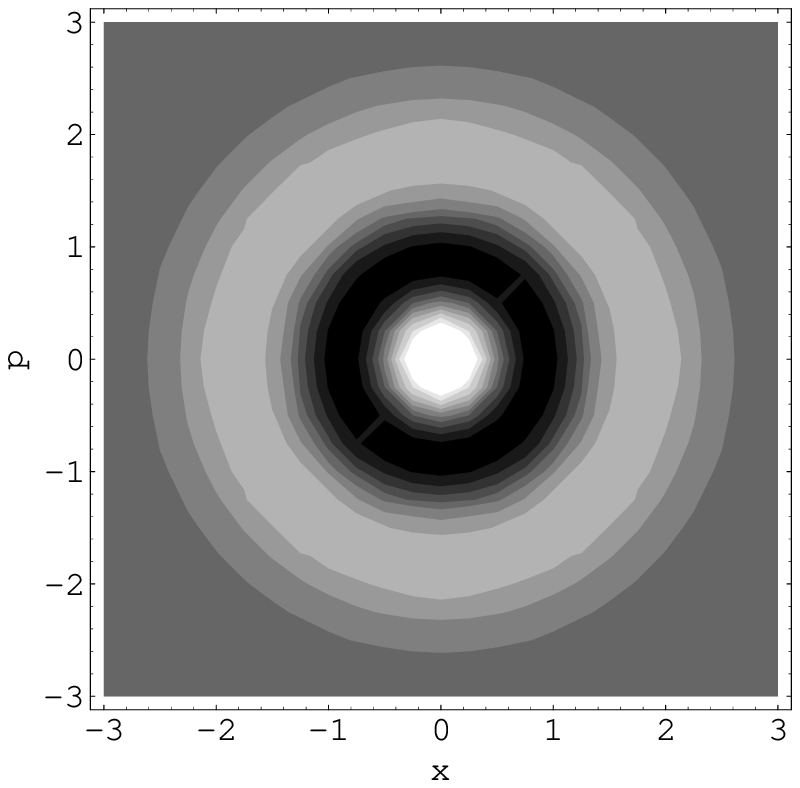}}
%\hspace{5mm}
\subfigure[\label{wignum2}]
{\includegraphics[width=6cm]{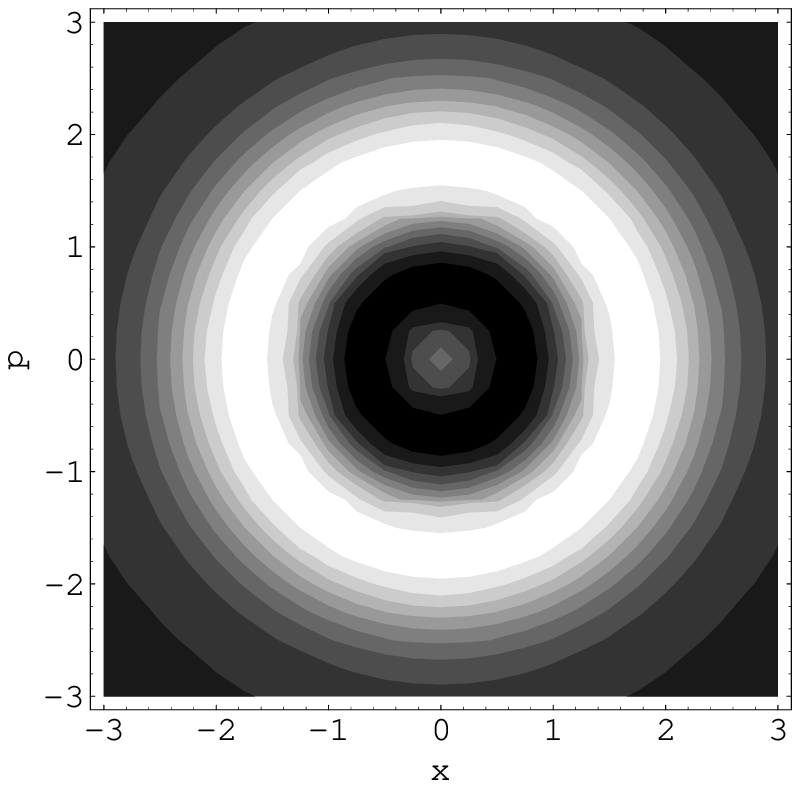}}
\subfigure[\label{wignum3}]
{\includegraphics[width=6cm]{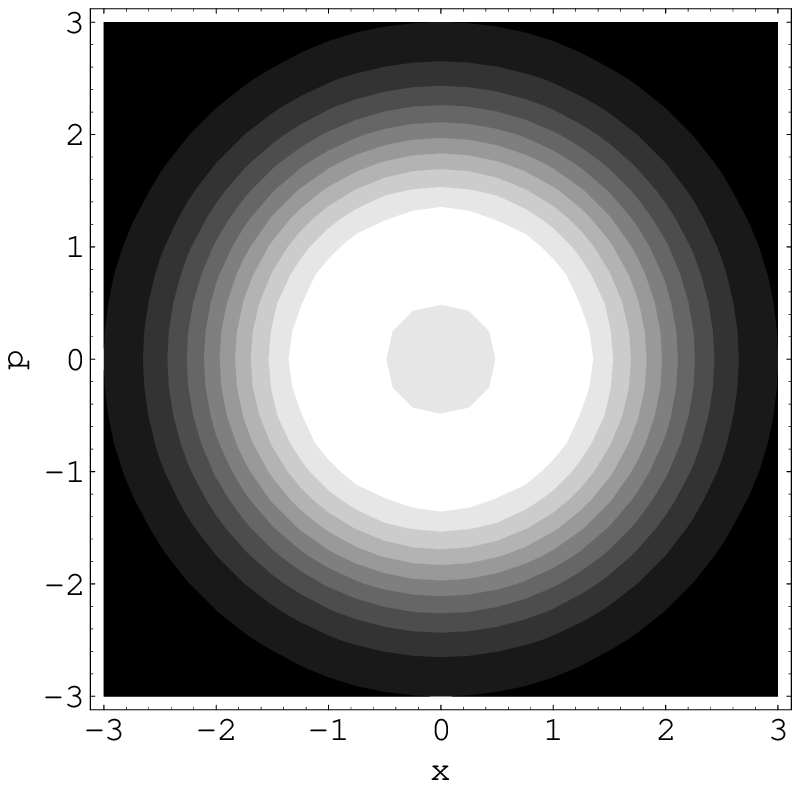}}
% \hspace{7mm}
\subfigure[\label{wignum4}]
{\includegraphics[width=6cm]{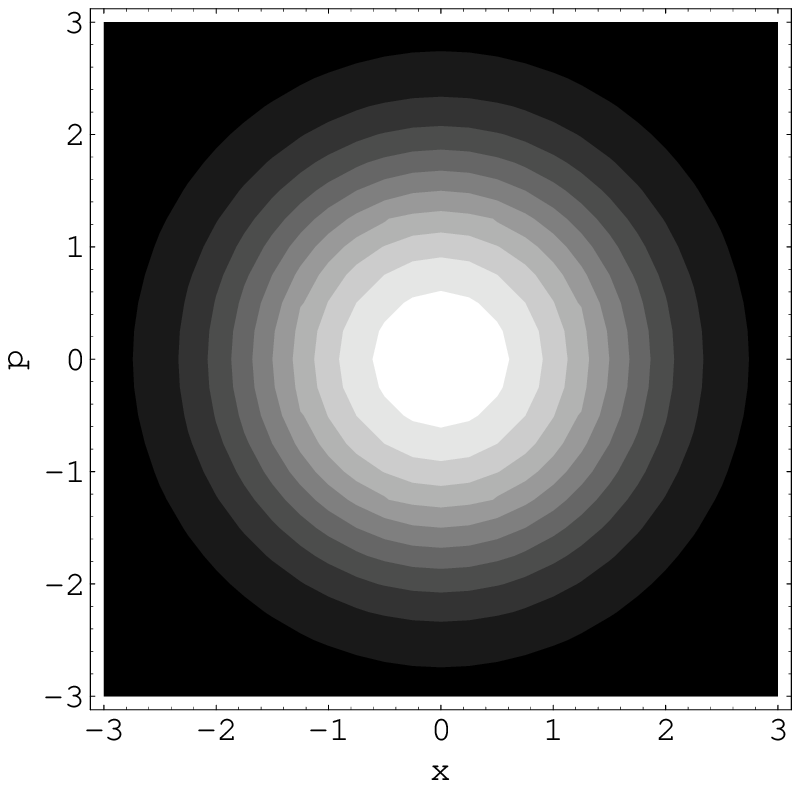}}
\end{center}
\caption{\sf \footnotesize Evolution in phase space of the Wigner function of the initial
number state $\ket{2}$  in a thermal channel with 
$\mu_{\infty}=0.5$ at times $t=0$ (a), $t=\gamma^{-1}/4$ (b), $t=\gamma^{-1}$ (c) 
and $t=1.5\gamma^{-1}$ (d). Darker colors stand for lower values, the scale of each plot 
is normalized. The time $t_{nc}$, at which the Wigner function of this state gets positive 
is $t_{nc}\simeq 0.4 \gamma^{-1}$. As can be seen, at $t=\gamma^{-1}$, the central
minimum deriving from the initial negative zone is still evident, but takes only positive 
values. 
\label{wignum}}
\end{figure*}
%%%%%%%%%%%%%%%%%%%%%%%%%%%%%%%%%%%%%%%%%%%%%%%%%%
%%%%%%%%%%%%%%%%%%%%%%%%%%%%%%%%%%%%%%%%%%%%%%%%%%
\begin{figure}[tb!]
\begin{center}
\includegraphics[scale=1]{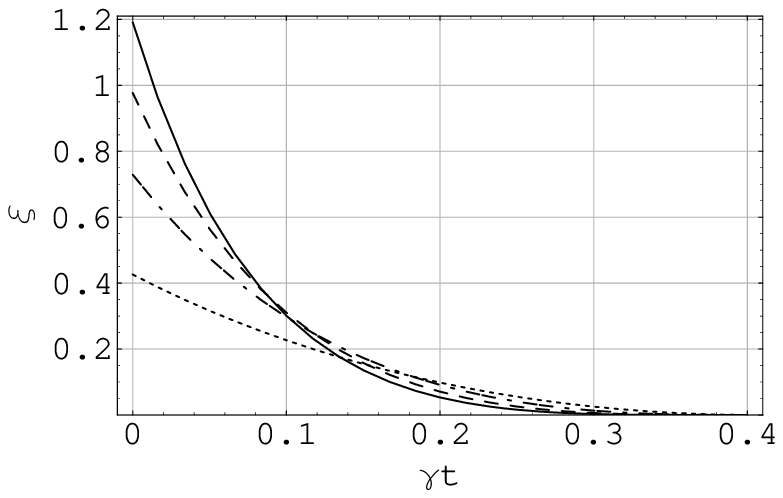}
\caption{\sf \footnotesize
Time evolution of the negative part $\xi$ of the number states $\ket{1}$ (dotted line),
$\ket{2}$ (dot-dashed line), $\ket{3}$ (dashed line) and $\ket{4}$ (continuous line),
in a thermal reservoir with $\mu_{\infty}=0.5$.
For such a reservoir, the Wigner function gets positive 
at $t_{nc}\simeq 0.4 \gamma^{-1}$. 
\label{numnc}}
\end{center}
\end{figure}
%%%%%%%%%%%%%%%%%%%%%%%%%%%%%%%%%%%%%%%%%%%%%%%%%%
The evolution in phase space 
of the Wigner function of \eq{wign} is showed in Fig.~\ref{wignum}. 
Fig.~\ref{numnc} shows the time dependence of the negative part $\xi$, numerically integrated 
for the first four number states in a thermal reservoir. Even though the initial negative part 
increases with increasing $n$, the quantity $\xi(t)$ is not increasing with $n$ at any time: indeed,
lower order states better preserve such nonclassical features when approaching the time
$t_{nc}$ (which, we recall once again, does not depend on the initial pure non Gaussian state).

\begin{figure}[tb]
\begin{center}
\includegraphics[scale=1]{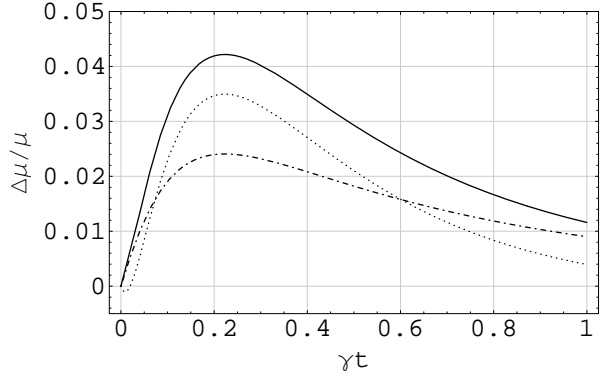}
\caption{\sf \footnotesize The relative increase in purity, defined by $\Delta\mu/\mu=
(\mu_{01}(t,r)-\mu_{01}(t,0))/\mu_{01}(t,0)$, as a function of time during the
evolution of the superposition $\ket{\psi_{01}}$ in Gaussian channels. 
The optimal condition $\vartheta=\varphi+\pi/2$ is always assumed, while $\mu_{\infty}=0.25$. 
The solid line refers to a bath with $r=0.28$, close to the optimal value;
the dotted line refers to a bath with $r=0.4$ and the dot--dashed line refers to a bath
with $r=0.1$.\label{catsnova}}
\end{center}
\end{figure}

A relevant instance to exemplify the 
decoherence of number states is
provided by the coherent normalized
superposition $\ket{\psi_{01}}=(\ket{0}+\,{\rm e}^{i\vartheta}\ket{1})/\sqrt{2}$,
constituting a microscopic Schr\"odinger cat.
The characteristic function $\chi_{01}$ of this state 
is simply found \cite{barnett}
\be
\chi_{01}(\alpha)=\frac{{\rm e}^{-\frac{|\alpha|^2}{2}}}{2}
\left[2-\,{\rm e}^{-\gamma t}|\alpha|^2-\,{\rm e}^{-\frac{\gamma t}{2}}
(\alpha^{*}\,{\rm e}^{-i\vartheta}-\alpha\,{\rm e}^{i\vartheta})\right]  .
\ee
Inserting $\chi_{01}$ as the initial condition in \eq{solu} and 
performing the integration of \eq{wigpur} yields, for the 
purity of the initial cat-like state evolving in the channel
\bea
\mu_{01}(t,r)&=&4\nu-\,{\rm e}^{-2\gamma t}\frac{\nu^2}{2\mu_{\infty}}
\Big(\mu_{\infty}+(\,{\rm e}^{\gamma t}-1)(\cosh(2r) \nonumber\\
&+&\cos(2\vartheta-2\varphi)\sinh(2r))\Big)\nonumber\\
&+&\,{\rm e}^{-4\gamma t}\frac{\nu^5}{2\mu_{\infty}^2}
\Big(4\mu_{\infty}^2+8(\,{\rm e}^{\gamma t}-1)\mu_{\infty}\cosh(2r)\nonumber\\
&+&(\,{\rm e}^{\gamma t}-1)^2(3\cosh(4r)+1)\Big)
\label{purcatn}\eea
where 
\be
\nu=\bigg[\frac{1}{\mu_{\infty}^{2}}\left(1-
{\rm e}^{-\gamma t}\right)^{2} \, + \,
{\rm e}^{-2\gamma t}
+2\frac{1}{\mu_{\infty}}\cosh(2r)
\bigg]^{-1/2}
\ee
is the purity of an initial vacuum in the channel, found in Sec.~\ref{1mode}. \eq{purcatn}
shows that the evolution of the coherent superposition 
is sensitive to the phase $\varphi$ of the bath. It is straightforward 
to see that the optimal choice maximizing purity at any given time 
is provided by $\vartheta=\varphi+\pi/2$. Fixing such a choice, 
we have numerically analyzed the dependence of $\mu_{01}$ on
the squeezing parameter $r_{\infty}$.
For small $r_{\infty}$ the purity $\mu_{01}$ increases with $r$. 
The optimal choice for $r_{\infty}$ depends on time, for $\gamma t=0.5$ it turns out to
be $r\simeq0.28$. The relative increase in purity for several choices of the
squeezing parameter $r_{\infty}$ is plotted
in Fig.~\ref{catsnova} as a function of time.
It is interesting to compare this analysis of decoherence with the one previously 
carried out for Gaussian catlike states.
Indeed, notwithstanding the deeply quantum nature of a superposition 
of number states, its decoherence rate is comparatively slow. Actually, the purity 
of the considered superposition in a thermal channel reaches the asymptotic value of the 
channel, after the initial decrease, in a time $t\simeq 0.5 \gamma^{-1}$. Such a time length 
corresponds to the decoherence time $t_{dec}$ of a superposition of two Gaussian terms 
displaced in phase space of only one coherent photon (in opposite directions 
with respect to the origin, {\em i.e.~}with $\|X_0\|^2=1$ in the notation of 
the previous section).
Despite the relevant intrinsic differences between these two kinds of
Schr\"odinger cat states, their decoherence is basically driven by the same process,
due to the entanglement of the system with the environmental degrees of freedom.

We remark that the time of decoherence can be much shorter of the time
characterizing the energy relaxation \cite{walls,myatt}, which constitutes however
a strict upper bound on the former.
This fact is a manifestation of a general feature of quantum mechanics.
Nonclassical superpositions decohere on a time scale of the photon lifetime in the channel,
regardless of the other parameters: once a single photon is added or lost, all the information
contained in the original state leaks out to the environment. This can be understood,
euristically, by considering the action of the annihilation operators $a$ which,
in general, modifies the coherent phase of the superposition. Therefore, as soon as the 
probability of losing a photon reaches $0.5$,  the original superposition turns into an
incoherent mixtures of states with different phase, whose interference terms cancel out
each other \cite{garraway,harochelh}.
No coherent behaviour can survive such a dissipative process and be 
afterwards revealed by interferometry.
%%%%%%%%%%%%%%%%%%%%%%%%%%%%%%%%%%%%%%%%%%%%%%%%%%
%%%%%%%%%%%%%%%%%%%%%%%%%%%%%%%%%%%%%%%%%%%%%%%%%%
%%%%%%%%%%%%%%%%%%%%%%%%%%%%%%%%%%%%%%%%%%%%%%%%%%

\section{Two-mode Gaussian states}\label{2mode}

Two-mode Gaussian states are the simplest example of continuous variable bipartite states. 
Their decoherence under the quantum optical master equation can be therefore characterized
also by investigating the evolution of the correlations between the two modes of the systems. 
In particular, the decay of {\em quantum} correlations, {i.e.~}of the entanglement,
quantified by the logarithmic negativity, may be adopted as an indicator of decoherence.
Due to their clear interest, concerning both applications in quantum information and 
the study of fundamental features of entanglement, the behaviour of two-mode 
Gaussian states under non unitary evolutions has attracted a remarkable theoretical 
interest in later years \cite{duan97,rajapra,hiroshima01,scheel01,paris02,kim03,prauz03,serafini04}.
We review here the results of Ref.~\cite{serafini04};
moreover, we consider the instance of different couplings to the bath and provide 
a detailed study of the evolving nonclassical depth. \par
Before addressing the analysis of their decoherence in detail, let us
recall some basic facts about two-mode Gaussian states.
The $4\times 4$ covariance matrix $\boldsymbol{\sigma}$ shall be conveniently
written in terms of the three $2\times 2$
submatrices $\boldsymbol{\alpha}$, $\boldsymbol{\beta}$, $\boldsymbol{\gamma}$
\begin{equation}
\boldsymbol{\sigma}\equiv\left(\begin{array}{cc}
\boldsymbol{\alpha}&\boldsymbol{\gamma}\\
\boldsymbol{\gamma}^{T}&\boldsymbol{\beta}
\end{array}\right)\, . \label{espre}
\end{equation}
The CM $\boldsymbol{\sigma}$ can be put into the so called standard form 
$\boldsymbol{\sigma}_{sf}$ through a
local symplectic operation $S_{l}=S_{1}\oplus S_{2}$
\begin{equation}
S_{l}^{T}\boldsymbol{\sigma}S_{l}=\boldsymbol{\sigma}_{sf}
\equiv \left(\begin{array}{cccc}
a&0&c_{1}&0\\
0&a&0&c_{2}\\
c_{1}&0&b&0\\
0&c_{2}&0&b
\end{array}\right)\; . \label{stform}
\end{equation}
In what follows, let us suppose $|c_2|\ge|c_1|$. 
States whose standard form fulfills $a=b$ are said to be symmetric.
Let us recall that any pure state is symmetric and fulfills 
$c_{1}=-c_{2}=\sqrt{a^2-1/4}$. 
The correlations $a$, $b$, $c_{1}$, and $c_{2}$ are determined by the four local symplectic 
invariants ${\rm Det}\boldsymbol{\sigma}=(ab-c_{1}^2)(ab-c_{2}^2)$, 
${\rm Det}\boldsymbol{\alpha}=a^2$, ${\rm Det}\boldsymbol{\beta}=b^2$, 
${\rm Det}\boldsymbol{\gamma}=c_{1}c_{2}$. Therefore, the standard form 
corresponding to any covariance matrix is unique
(up to a common sign flip in the $c_i$'s).  \par
The $Sp_{(4,{\mathbb R})}$ invariants ${\rm Det}\gr{\sigma}$ and
$\Delta(\gr{\sigma})={\rm Det}\boldsymbol{\alpha}+\,{\rm Det}\boldsymbol{\beta}+2
\,{\rm Det}\boldsymbol{\gamma}$ permit to explicitly express Ineq.~(\ref{unce}) in terms of 
second moments
\begin{equation}
\Delta(\gr{\sigma})\le\frac{1}{4}+4\,{\rm Det}\boldsymbol{\sigma}
\label{sepcomp}\;
\end{equation}
and determine the symplectic spectrum $\{\nu_{\mp}\}$ of $\gr{\sigma}$,
according to \cite{serafozzi}
\[
2\nu_{\mp}^2 = \Delta(\gr{\sigma})\mp\sqrt{\Delta(\gr{\sigma})^2-4\,{\rm Det}\,
\gr{\sigma}} \; .
\]

A relevant subclass of Gaussian states we will make use of is 
constituted by the two--mode squeezed thermal states. Let 
$S_{r}=S_{12,r,0}$ 
be the two-mode squeezing operator between the modes $1$ and $2$
with real squeezing parameter $r$ and let $\nu_{\mu}$
be the tensor product of identical thermal states of global purity $\mu$,
with CM $\gr{\nu}_{\mu}=1/(2\sqrt{\mu}){\mathbbm 1}$.
Then, for a two-mode squeezed thermal state ${\xi}_{\mu,r}$
we can write ${\xi}_{\mu,r}=S_{r}{\nu}_{\mu}S_{r}^{\dag}$.
The CM $\gr{\xi}_{\mu,r}$ of $\xi_{\mu,r}$ is a symmetric standard form
satisfying
\be
a=\frac{\cosh2r}{2\sqrt{\mu}}\; , \quad 
c_{1}=-c_{2}=\frac{\sinh2r}{2\sqrt{\mu}} \, . \label{sqthe}
\ee
In the instance $\mu=1$ one recovers the pure
two--mode squeezed vacuum states. 
Two--mode squeezed states are endowed with remarkable properties 
related to entanglement \cite{2max}, in particular they are the maximally 
entangled states for given marginal and global purities \cite{adesso,adesso2}.

We recall that the necessary and sufficient separability criterion for
two-mode Gaussian states is  
positivity of the partially transposed density matrix 
(``PPT criterion'') \cite{simon00}. 
It can be easily
seen from the definition of $W(X)$ that the action of partial transposition 
amounts, in phase space, to a mirror reflection of one of the four canonical variables. 
In terms of the $Sp_{4,\mathbb{R}}$ invariants,
this results in changing
the invariant $\Delta(\gr{\sigma})$ into $\tilde{\Delta}({\gr\sigma})
=\Delta(\tilde{\gr{\sigma}})=\,{\rm Det}\,\gr{\alpha}+
\,{\rm Det}\,\gr{\beta}-2\,{\rm Det}\,\gr{\gamma}$. Now, the symplectic
eigenvalues $\tilde{\nu}_{\mp}$ of the partially transposed CM $\tilde{\gr{\sigma}}$ read
\be
\tilde{\nu}_{\mp}=
\sqrt{\frac{\tilde{\Delta}(\gr{\sigma})\mp\sqrt{\tilde{\Delta}(\gr{\sigma})^2
-4\,{\rm Det}\,\gr{\sigma}}}{2}} \, . \label{sympareig}
\ee
The PPT criterion then reduces to a simple inequality that must
be satisfied by the smallest symplectic eigenvalue $\tilde{\nu}_{-}$
of the partially transposed state
\be
\tilde{\nu}_{-}\ge \frac12 \: ,
\label{symppt}
\ee
which is equivalent to 
\be
\tilde{\Delta}(\gr{\sigma})\le 4\,{\rm Det}\,\gr{\sigma}+\frac14 \; .
\label{ppt}
\ee 
The above inequalities
imply ${\rm Det}\,\gr{\gamma}=c_{1}c_{2}<0$ 
as a necessary condition for a two--mode Gaussian state 
to be entangled.
The quantity $\tilde{\nu}_{-}$ encodes all the qualitative characterization of 
the entanglement for arbitrary (pure or mixed) two--modes Gaussian states. 
Note that
$\tilde{\nu}_{-}$ takes a particularly simple form for 
entangled symmetric states, whose standard 
form has $a=b$
\be
\tilde{\nu}_{-}=\sqrt{(a-|c_{1}|)(a-|c_{2}|)} \; .
\label{symeig}
\ee
The logarithmic negativity $E_{\N}$ of two--mode Gaussian states  
is a simple function of $\tilde{\nu}_{-}$, which is
thus itself an (increasing) entanglement monotone;
one has in fact \cite{adesso2}
\be
E_{\N}(\gr{\sigma})=\max\left\{0,-\ln{2\tilde{\nu}_{-}}\right\} \: .
\ee
This is a decreasing function of the smallest partially transposed symplectic 
eigenvalue $\tilde{\nu}_{-}$, quantifying the amount by which Inequality 
(\ref{symppt}) is violated. Thus, for our aims, 
the eigenvalue $\tilde{\nu}_{-}$ completely qualifies and quantifies
the quantum entanglement of a two--mode Gaussian state $\gr{\sigma}$.\par

The smallest eigenvalue $u$ of $\gr{\sigma}_{sf}$ (which determines the nonclassical 
depth $\tau$ according to \eq{ncgau}) is easily determined 
\be
2u = a+b - \sqrt{(a-b)^2 + 4 c_2^2} \; ,
\label{u2m}
\ee
reducing to $u=a-|c_2|$ for symmetric states and to $u=\,{\rm e}^{-2r}/(2\sqrt{\mu})$
for two-mode squeezed thermal states.

The evolution of two-mode Gaussian states in the noisy channel is described 
by \eq{2evo} with $n=2$. The channel is completely determined by the quantities 
$\mu_{i\infty}$, $r_{i\infty}$, $\varphi_{i\infty}$ and $\gamma_{i}$, for
$i=1,2$. Notice that, if $\gamma_1\neq \gamma_2$, then a change in the values of the 
couplings to the bath $\gamma_i$'s does not reduce to a rescaling of time and 
may significantly affect
the evolution of the relevant quantities in the channel.
For the study of the entropic measures and of correlations, we will restrict to initial 
states in the standard form of \eq{stform}, with no loss of generality since 
all such quantities are invariant under local unitary operations. On the other hand, 
the nonclassical depth $\tau$ is not invariant under such operations. 
Determining the evolution of such a quantity in the general instance is slightly more 
involved. For the sake of simplicity, we will study such evolution in relevant instances,
which can be conveniently handled and illustrate the general behaviour of the nonclassical indicator.
Henceforth, we will set $\varphi_{1\infty}=0$ as a reference choice for phase space rotations.

Exploiting the results we have just reviewed, together with the general
definitions of Sec.~\ref{nota}, we can determine the exact evolution in
the channel of the entropic measures $\mu$ and $S_V$, and of the quantum
and total correlations, respectively
quantified by $E_{\cal N}$ and $I$. In Appendix \ref{coeffi} we provide
the explicit expression of the 
time dependent terms, allowing to compute such evolutions, in the  
instance of equal couplings: $\gamma_1=\gamma_2=\gamma$.  

\begin{figure}[tb!]
\begin{center}
\includegraphics[scale=1]{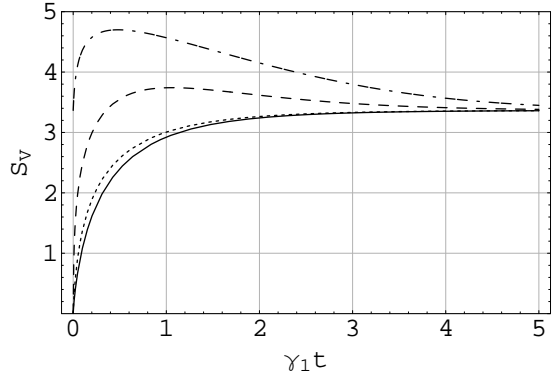}
\caption{\sf \footnotesize
Time evolution of the von Neumann entropy in a thermal channel with
$\gamma_1=1$, $\gamma_2=2$ and $\mu_{1\infty}=\mu_{2\infty}=0.25$.
The continuous line refers to the conjectured optimal evolution, achieved by
a pure separable input state with $a=b=1/2$ and $c_1=c_2=0$ in a non 
squeezed channel; the dotted line refers to a n inital pure two mode squeezed state 
with $r=0.5$ in the same channel. The dashed and dot-dashed lines refer to
a squeezed channel with $r_{1\infty}=r_{2\infty}=1$ and, respectively to an 
initial pure two mode squeezed state with $r=1$ and an intial thermal two mode 
squeezed state with $\mu=1/16$ (equal to the asymptotic purity).
\label{multi}}
\end{center}
\end{figure}

As for the evolution of the purity $\mu$ and of the von Neumann entropy $S_V$
-- whose decrease quantifies the information which the composite two mode
state `as a whole' loses by interacting with the environment -- some
analytical statements can be done. It can be shown by means of a variational 
approach \cite{canali} that the purity of a given channel of the form of \eq{2evo} is maximized
by an uncorrelated state (with $c_1=c_2=0$ in our notation). 
Its maximization is therefore achieved by the 
(obviously separable) product of two `countersqueezed' states, which, as we
have seen in Sec.~\ref{1mode} maximizes the local purity relative to the 
two single-mode channels.\footnote{This is a particular instance in which, 
rstricting to the Gaussian setting, the maximal output purity of a 
tensor product of channels is `multiplicative' \cite{canali}.}
The optimal purity evolution reduces therefore to the square of the optimal
purity evolution for single mode channels, previously studied. This feature 
holds for any value of $\gamma_1$ and $\gamma_2$. An analogous argument can
be applied to the von Neumann entropy $S_V$ which, we recall, is fully 
determined by the quantity $\lim_{p\rightarrow 0}\,{\rm Tr}\,\varrho^{p}$. 
However, so far, the fact that the minimal $S_V$ at any given time
is achieved by an uncorrelated input has been proved only for 
$\gamma_1=\gamma_2$. The numerical analysis, summarized in Fig.~\ref{multi}, 
remarkably supports the conjecture of the additivity of the minimal output 
von Neumann entropy also for $\gamma_1\neq\gamma_2$.\footnote{The {\em 
additivity} of the minimal von Neumann entropy corresponds to the 
{\em multiplicativity} of the maximum of the quantity 
$\lim_{p\rightarrow 0}\,{\rm Tr}\,\varrho^{p}$.}
%%%%%%%%%%%%%%%%%%%%%%%%%%%%%%%%%%%%%%%%%%%%%%%%%%
%%%%%%%%%%%%%%%%%%%%%%%%%%%%%%%%%%%%%%%%%%%%%%%%%%

We now move to consider the decay of the entanglement 
between the two modes of the field, {\em i.e.~}the leaking to the 
environment of the information contained in quantum correlations between 
the two modes. 
Supposing that the couplings to the two baths are equal
($\gamma_1=\gamma_2=\gamma$) and
making use of the separability criterion given by Ineq.~(\ref{ppt}), 
one finds that an initially entangled state becomes separable 
at a certain time $t$ if
\begin{equation}
u\,{\rm e}^{-4\gamma t}+v\,{\rm e}^{-3\gamma t}+w\,{\rm e}^{-2\gamma t}+
y\,{\rm e}^{-\gamma t}+z=0 \, . \label{dis4}
\end{equation}
The coefficients $u$, $v$, $w$, $y$ and $z$ are functions of the nine parameters 
characterizing the initial state and the channel (see App.~\ref{coeffi}).\footnote{Clearly,
in the general instance of different couplings ($\gamma_1\neq\gamma_2$),
\eq{dis4} would turn in a system of fourth degree in the two unlnown
$\,{\rm e}^{-\gamma_1 t}$ and $\,{\rm e}^{-\gamma_2 t}$. Such a situation does not
pose any conceptual problem and can be treated in much the same way as the one here
described, by explicitly determining the coefficients of the system.} 
Eq.~(\ref{dis4}) is an algebraic equation of fourth degree
in the unknown $k={\rm e}^{-\gamma t}$. 
The solution $k_{ent}$ of such an equation  
closest to one, and satisfying $k_{ent}\le 1$
can be found for any given initial
entangled state. Its knowledge promptly leads to the determination 
of the ``entanglement time'' $t_{ent}$ of
the initial state in the channel, defined as the time interval after which the initial 
entangled state 
becomes separable
\begin{equation}
t_{ent}=-\frac1\gamma \ln k_{ent} \, .
\label{loga}
\end{equation} \par
The entanglement time $t_{ent}$ can be easily estimated for
symmetric states (for which $a=b$) evolving in equal thermal baths ({\em i.e.~}with 
$\gamma_1=\gamma_2=\gamma$ and $\mu_{1\infty}=\mu_{2\infty}=\sqrt{\mu_{\infty}}$).
In such a case the initally entangled state maintains its symmetric standard form 
during the time evolution.
Recalling that $|c_{1}|\le|c_{2}|$, we have that 
Eqs.~(\ref{symppt}) and (\ref{symeig}) provide
the following bounds for the entanglement time
\be
\ln\left(1+
\sqrt{\mu_{\infty}}
\frac{2|c_{1}|-2a+1}{1-\sqrt{\mu_\infty}}\right)\le\gamma t_{ent}\le
\ln\left(1+
\sqrt{\mu_\infty}
\frac{2|c_{2}|-2a+1}{1-\sqrt{\mu_\infty}}\right) \, . \label{tentbnd}
\ee
Note that $\mu_{\infty}$ is the global purity of the asymptotic two mode state.
Imposing the additional property $c_{1}=-c_{2}$
amounts to consider standard forms which can be written as squeezed thermal states
(see Eqs.~\ref{sqthe}). 
For such states, Inequality (\ref{tentbnd}) reduces to
\be
t_{ent}=\frac{1}{\gamma}\ln\left(1+
\sqrt{\mu_{\infty}}\frac{1-\frac{{\rm e}^{-2r}}{\sqrt{\mu}}}
{1-\sqrt{\mu_\infty}}\right)\; . \label{tenteq}
\ee
In particular, for $\mu=1$, one recovers the entanglement 
time of a two--mode squeezed vacuum state in a thermal channel
\cite{duan00,paris02, prauz03}. We point out that two--mode
squeezed vacuum states encompass all the possible standard forms of 
pure Gaussian states.

\begin{figure}[tb!]
\begin{center}
\includegraphics[scale=1]{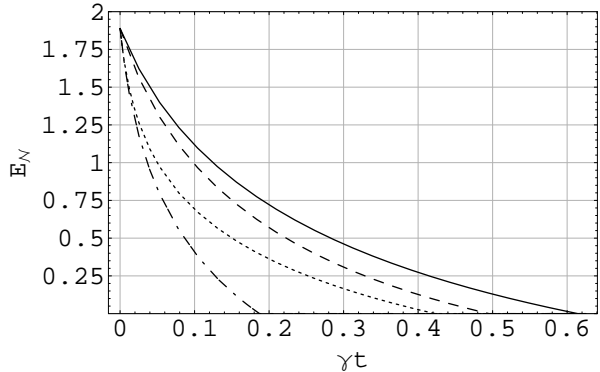}
\caption{\sf \footnotesize
Time evolution of the logarithmic negativity of an initial two mode squeezed thermal 
state with $\mu=0.8$ and $r=1$ in several channels with $\gamma_1=\gamma_2=\gamma$. 
The continuous line refers to a non squeezed bath with $\mu_{1\infty}=\mu_{2\infty}=
0.5$ (corresponding to $0.5$ thermal photons); the dotted line correspond to a thermal bath 
with $\mu_{1\infty}=0.25$ and $\mu_{2\infty}=1$ (with a global asymptotic purity
equal to the previous one); the dashed and dot-dashed lines refer to a bath with
$\mu_{1\infty}=\mu_{2\infty}=0.5$, $r_{1\infty}=r_{2\infty}=1$ and
$\varphi_{2\infty}=0$ ($\varphi_{2\infty}=\pi/4$) for the dashed (dot-dashed) line.  
\label{sqthlneg}}
\end{center}
\end{figure}

\begin{figure}[tb!]
\begin{center}
\includegraphics[scale=1]{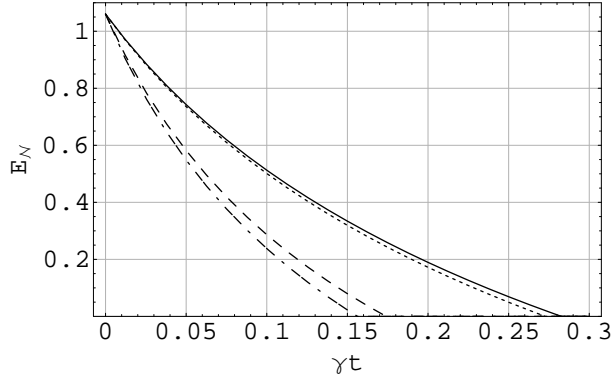}
\caption{\sf \footnotesize
Time evolution of the logarithmic negativity of an inital entangled non symmetric
state, obtained from the squeezed thermal one considered in Fig.~\ref{sqthlneg} by adding
$0.2$ to the element $a$ of the standard form (added noise on mode $1$ quadratures.) 
The solid line refers to a bath with $\gamma_1=\gamma_2=\mu_{1\infty}=
\mu_{2\infty}=1$; the dotted line refers to a channel with 
$\gamma_1=0.5$, $\gamma_2=1.5$ and $\mu_{1\infty}=
\mu_{2\infty}=1$; the dashed (dot-dashed) line refers to a bath with 
$\gamma_1=\gamma_{2}=1$, $\mu_{1\infty}=1/9$ ($\mu_{1\infty}=1$) and
$\mu_{2\infty}=1$ ($\mu_{2\infty}=1/9$). The label $\gamma$ is defined by
$\gamma=(\gamma_1+\gamma_2)/2$.  
\label{asymlneg}}
\end{center}
\end{figure}

The results of the numerical analysis of the evolution of the logarithmic negativity
for several initial states are reported 
in Figs.~\ref{sqthlneg} and \ref{asymlneg}.
In general, one can see that a less mixed environment better preserves 
entanglement 
by prolonging the entanglement time. More remarkably,   
Fig.~\ref{sqthlneg} shows that a local squeezing of the two 
uncorrelated channels
does not help to preserve the quantum correlations between the evolving modes. Moreover,
as can be seen from Fig.~\ref{asymlneg}, 
states with greater uncertainties on, say, mode $1$ ($a>b$) 
better preserves its entanglement if bath $1$ is more mixed 
than bath $2$ ($\mu_{1\infty}<\mu_{2\infty}$). Fig.~\ref{asymlneg} also shows 
that, even for initial non symmetric states,
unbalancing the couplings to the two single mode reservoirs (while leaving
their average unchanged: $\gamma=(\gamma_1+\gamma_2)/2$)
only slightly affects the evolution of the entanglement in the channel; 
an accurate numerical analysis shows that a greater 
coupling to the more mixed initial mode ({\em e.g.}, $\gamma_1>\gamma_2$ if 
$a>b$) enhances the preservation of the initial quantum correlations.
Also, for symmetric states evolving in squeezed baths, one can see that 
the entanglement of the initial state is better preserved if the
squeezing of the two channels is balanced.\par 
%%%%%%%%%%%%%%%%%%%%%%%%%%%%%%%%%%%%%%%%%%%%%%%%%%
%%%%%%%%%%%%%%%%%%%%%%%%%%%%%%%%%%%%%%%%%%%%%%%%%%
%%%%%%%%%%%%%%%%%%%%%%%%%%%%%%%%%%%%%%%%%%%%%%%%%%
%%%%%%%%%%%%%%%%%%%%%%%%%%%%%%%%%%%%%%%%%%%%%%%%%%

\begin{figure}[tb!]
\begin{center}
\includegraphics[scale=1]{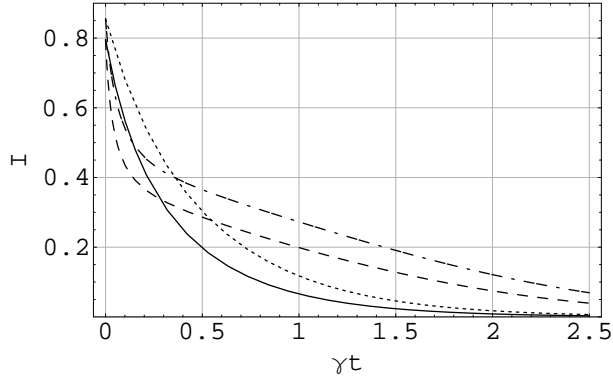}
\caption{\sf \footnotesize
Time evolution of the mutual information of Gaussian states
in an environment with $\gamma_1=\gamma_2=\gamma$ and
$\mu_{1}=\mu_{2}=1/3$. The continuous line refers to
an entangled state
with $a=2$, $b=c_{1}=-c_{2}=1$ in a non squeezed environment;
the dotted line refers to the same state
in an environment with $r_{1}=r_{2}=1$; the dashed line refers to a 
non entangled state 
with $a=b=2$, $c_{1}=-c_{2}=1.5$ in a non squeezed environment; the dot--dashed line refers 
to the same state in a squeezed environment with $r_{1}=r_{2}=1$. The squeezing angle 
$\varphi_{2}$ has always been set to $0$.
\label{minf}}
\end{center}
\end{figure}

An interesting feature concerns the evolution of the
mutual information $I$, illustrated in Fig.~\ref{minf} for some relevant cases:
at long times, such a quantity is better preserved in squeezed channels. This 
property has been thoroughly tested both on non entangled states, featuring 
only classical correlations, and on highly entangled states, and  seems to
hold generally.
%%%%%%%%%%%%%%%%%%%%%%%%%%%%%%%%%%%%%%%%%%%%%%%%%%
%%%%%%%%%%%%%%%%%%%%%%%%%%%%%%%%%%%%%%%%%%%%%%%%%%

The instance of a standard form state in a tensor product of two 
thermal channels (parametrized by $\gamma_i$ and $\mu_{i\infty}$, for $i=1,2$)
is especially relevant, since it gives a basic description of dissipation 
in most experimental settings, like fiber--mediated communication protocols.
A simple analysis straightforwardly shows that in this instance both the purity
and the logarithmic negativity (that is, the entanglement) of the evolving state
are increasing functions of the asymptotic purities and decreasing functions of the couplings to
the baths. This should be expected, recalling the well understood 
synergy between entanglement and purity for general quantum states: the ideal vacuum
environment, whose decoherent action is entirely due to losses, is the one which better 
preserve both the global information of a state and its correlations.\par
As we have seen,
two--mode squeezed thermal states constitute 
a relevant class of Gaussian states, parametrized 
by their purity $\mu$ and by the squeezing parameter $r$
according to Eqs.~(\ref{sqthe}).
In particular, 
two--mode squeezed vacuum states
(or twin-beams), which can be defined as
squeezed thermal states with $\mu=1$,
correspond to maximally
entangled symmetric states for fixed marginal purities \cite{adesso2}.
Therefore, they constitute a crucial resource
for quantum
information processing in the continuous variable scenario.
For squeezed thermal states (chosen as initial conditions
in the channel), it can be shown analytically that the
partially transposed symplectic eigenvalue $\tilde{\nu}_{-}$
is at any time an increasing function of the bath squeezing angle
$\varphi_{2}$: ``parallel'' squeezing in the two channels
optimizes the preservation of entanglement.
Both in the instance of two equal squeezed baths ({\it i.e.~}with
$r_{1}=r_{2}=r$) and of a thermal bath joined to
a squeezed one ({\it i.e.~}$r_{1}=r$ and $r_{2}=0$),
it can be shown that $\tilde{\nu}_{-}$ is an
increasing function of $r$ \cite{serafini04}.
Such analytical considerations, supported by a broader 
numerical analysis, clearly show that
a local squeezing of the environment faster
degrades the entanglement of the initial state. The same behavior 
occurs for purity.

\begin{figure}[tb!]
\begin{center}
\includegraphics[scale=1]{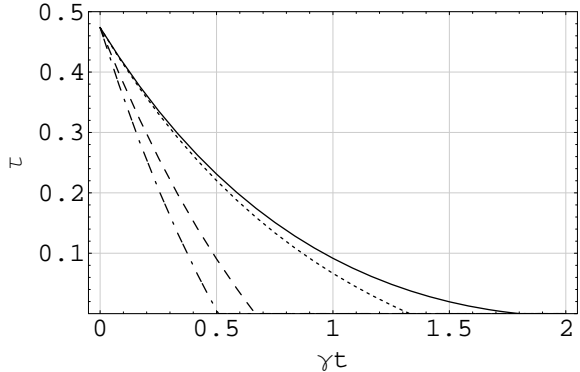}
\caption{\sf \footnotesize Evolution of the nonclassical depth of an initial two-mode 
squeezed thermal state with $\mu=0.9$ and $r=1.5$. The dashed line refers to the evolution 
in a non squeezed bath with $\mu_{1\infty}=\mu_{2\infty}=0.5$; the dot-dashed line
refers to a non squeezed bath with $\mu_{1\infty}=0.25$ and $\mu_{2\infty}=1$; 
the dotted line refers to a bath with $\mu_{1\infty}=\mu_{2\infty}=0.5$ and $r_1=
r_2=0.2$; finally, the continuous line refers to a bath with 
$\mu_{1\infty}=\mu_{2\infty}=0.5$, $r_{1\infty=0.6}$ and $r_{2\infty}=0$.
\label{2mnc}}
\end{center}
\end{figure}

In order to illustrate the behaviour of the nonclassical depth $\tau$ in the
noisy channel, let us consider standard form states evolving in thermal environments.
For simplicity, let us assume $\gamma_1=\gamma_2=\gamma$.
According to Eqs.~(\ref{ncgau}) and (\ref{u2m}), one has, for the evolving 
nonclassicality (recalling that $|c_2|\ge |c_1|$) 
\be
\begin{split}
\tau(t)=&\frac12-\frac12(a+b)\,{\rm e}^{-\gamma t}-
\frac{\mu_{1\infty}+\mu_{2\infty}}{4\mu_{1\infty}\mu_{2\infty}}
(1-\,{\rm e}^{-\gamma t})\\
&+\frac12\sqrt{\left((a-b)\,{\rm e}^{-\gamma t}+
\frac{\mu_{1\infty}-\mu_{2\infty}}{2\mu_{1\infty}\mu_{2\infty}}
(1-\,{\rm e}^{-\gamma t})\right)^2+4c_2^2 \,{\rm e}^{-2\gamma t}}\Bigg) \, .
\end{split}
\ee
This function is a decreasing function of the parameters $\mu_{i\infty}$: 
the thermal noise contributes to destroy the nonclassical features of the initial state. 
To study the effect of the squeezing of the bath on the nonclassical depth, we specialize
to the instance of two mode squeezed thermal states, which are an archetypical class of
nonclassical two mode states, characterized by squeezing in combined quadratures.
In this case 
it can be easily shown that, in order to minimize the smaller eigenvalue of $\gr{\sigma}$ 
(thus maximizing $\tau$) the choice $\varphi_2 = 0$ is optimal. 
We will thus make such a choice in the following. The nonclassical depth of the initial 
two mode squeezed state $\xi_{\mu,r}$ in a channel with parameters $\mu_{i\infty}$
and $r_{i\infty}$ for $i=1,2$, takes the following form
\be
\begin{split}
\tau(t) =& \frac12 - \frac{\cosh(2r)}{2\sqrt{\mu}}\,{\rm e}^{-\gamma t} - 
\frac{\,{\rm e}^{-2r_{1\infty}}\mu_{2\infty}+\,{\rm e}^{-2r_{2\infty}}\mu_{1\infty}}
{4\mu_{1\infty}\mu_{2\infty}}(1-\,{\rm e}^{-\gamma t})\\
&+\frac12\sqrt{\left(\frac{\,{\rm e}^{-2r_{1\infty}}\mu_{2\infty}-
\,{\rm e}^{-2r_{2\infty}}\mu_{1\infty}}
{2\mu_{1\infty}\mu_{2\infty}}(1-\,{\rm e}^{-\gamma t})\right)^2+
\frac{\sinh(2r)^2}{\mu}\,{\rm e}^{-2\gamma t}}\, .
\end{split}\label{2msqnc}
\ee
\eq{2msqnc} reduces to the following simple form for the evolution in equal baths
(with $\mu_{1\infty}=\mu_{2\infty}=\sqrt{\mu}$ and $r_{1\infty}=
r_{2\infty}=r_{\infty}$)
\be
\tau(t)= \frac{1-\frac{{\rm e}^{-2r}}{\sqrt{\mu}}\,{\rm e}^{-\gamma t}
-\frac{{\rm e}^{-2r_\infty}}{\sqrt{\mu_\infty}}\,(1-{\rm e}^{-\gamma t})}{2} \; .
\ee
As can be seen in Fig.~\ref{2mnc}, the local squeezing of the baths, reducing 
the quantum noise in one quadrature of the multimode system, 
drastically increases the duration of the nonclassicality of the state and, generally, 
the value of its nonclassical depth at any given time. This is due to the symmetry
of two mode squeezed states under mode exchange: such states can take advantage
of reduced fluctuations of {\em any} quadrature of the bath. Interestingly, while 
the nonclassical depth is enhanced by the local squeezing of a quadrature (thus 
implying an improved preservation of nonclassical features like subpoissonian 
photon number distributions), the entanglement is not. This is due to the intrinsically
non local nature of the entanglement: the advantage which could be achieved by squeezing
a local quadrature is balanced by the increased fluctuations in the conjugated quadrature, 
which usually makes squeezing not favourable to the aim of preserving entanglement. 
%%%%%%%%%%%%%%%%%%%%%%%%%%%%%%%%%%%%%%%%%%%%%%%%%%
%%%%%%%%%%%%%%%%%%%%%%%%%%%%%%%%%%%%%%%%%%%%%%%%%%

\section{Concluding remarks}\label{concl}

We have carried out a quantitative analysis of decoherence of continuous variable systems
interacting with general Gaussian environments and reviewed
many related results. The method we have presented
to study the decoherence rate may be
applied to other systems of interest, like qubit systems under non unitary evolutions.
Several relevant configurations have been considered and exhaustively analysed, characterizing
their rate of decoherence by keeping track of the decay of the global
degree of purity,
of indicators of nonclassicality and, for two mode states, of
quantum and total correlations.

Quite in general, we have shown that, as long as one restricts to the Gaussian setting,
squeezing the bath (or, equivalently, the intial state while letting the bath being thermal)
does not help to better preserve either the overall coherence of the state or its
quantum correlations. However, such a squeezing proves effective in delaying the
decoherence of more deeply non classical states, like cat-like states resulting from
coherent superpositions of Gaussian states or of number states.
Furthermore, quite interestingly,
we have shown that a local squeezing of the baths may improve
the preservation of the mutual information in two-mode systems.

We remark that our results are of direct interest to recent
developments in experimental quantum optics, especially related 
to quantum information and quantum control.
Indeed, a crucial step towards the development of quantum information
technology is the achievement of a sufficent quantum control
capability, {\em i.e.~}of the ability of engineering quantum signals
and feedback techniques acting on
the dynamics of a quantum system. 
In fact, the implementation of any quantum information protocol relies on
maintaining quantum coherence in the system for a significant period of time
and so requires some kind of mechanism to eliminate or
mitigate the undesirable effects of decoherence.                                                                        
In this framework, a precise knowledge of the decoherence
dynamics is desirable, especially in the continuous variable 
regime, where
the field of quantum control originated and has a strong
experimental impact \cite{wieman99,vandersypen04}.

In order to make this point clearer, let us explicitly 
regard the following example. 
Consider the continuous variable teleportation of a single-mode 
coherent state by exploiting a two-mode squeezed thermal state 
as entangled resource (for a detailed description of the 
protocol, see Ref.~\cite{braunstein98}). Now, it may be shown \cite{adillu2}
that the optimal teleportation fidelity $F$
(averaged over the whole complex plane)
for such a protocol is given by a simple function of the smallest 
partially transposed symplectic eigenvalue $\tilde{\nu}_-$
of the two--mode squeezed state:
\be
F=\frac{1}{1+2\tilde\nu_{-}} \; . \label{telefid}
\ee
If the two modes which share the entangled state are, say, stored 
in two distant cavities, waiting to be used, the decoherence they experience
will gradually corrupt the fidelity of the teleportation protocol. 
Our study allows to keep track of the quantity $\tilde{\nu}_{-}$ during 
the dissipative evolution of the state as a function of various environmental 
parameters, and thus to exactly determine the teleportation fidelity achievable
as a function of time. For instance, considering an initially pure 
shared two-mode squeezed vacuum with squeezing parameter $r$, evolving in 
two environments with, for simplicity, the same coupling $\gamma$
and asymptotic purity $\mu_{\infty}$,
one gets
\be
F(t) = \frac{1}{1+\,{\rm e}^{-2r-\gamma t}+(1-\,{\rm e}^{-\gamma t})/\mu_{\infty}} \, .
\ee
Notice that such a result takes into account both losses and thermal noise.
In the more general instance, let us remark that the entanglement time, 
extensively analyzed in Sec.~\ref{2mode} [see Eqs.~(\ref{loga}-\ref{tenteq})] 
and which may be analytically 
determined following the approach we have presented, 
coincides with the time over which quantum teleportation allows to beat 
the classical fidelity, equal to $0.5$, as shown by \eq{telefid} (at $t_{ent}$
$\tilde\nu_{-}$ reaches 1/2 and then keeps increasing). 
After such a time the entanglement is gone 
because of local decoherence: the shared resource becomes useless 
to quantum informational aims.

%%%%%%%%%%%%%%%%%%%%%%%%%%%%%%%%%%%%%%%%%%%%%%%%%%

\appendix
\section{Determination of mixedness and entanglement of two-mode states
\label{coeffi}}
Here we provide explicit expressions which allow to determine the exact evolution
in uncorrelated channels with $\gamma_1=\gamma_2=\gamma$ 
of a generic initial state in standard form. 
The relevant quantities $E_{\N}$, $\mu$, $S_{V}$, $I$ and $\tau$ 
are all functions of the four $Sp_{(2,\mathbb{R})}\oplus Sp_{(2,\mathbb{R})}$
invariants $\,{\rm Det}\,\gr{\alpha}$, $\,{\rm Det}\,\gr{\beta}$
$\,{\rm Det}\,\gr{\gamma}$ and $\,{\rm Det}\,\gr{\sigma}$. 
Let us then write these quantities as follows
\bea
{\rm Det}\,\gr{\sigma}&=&\sum_{k=0}^{4}\Sigma_{k}\,{\rm e}^{-k\Gamma t}\; ,\\
{\rm Det}\,\gr{\alpha}&=&\sum_{k=0}^{2}\alpha_{k}\,{\rm e}^{-k\Gamma t}\; ,\\
{\rm Det}\,\gr{\beta}&=&\sum_{k=0}^{2}\beta_{k}\,{\rm e}^{-k\Gamma t}\; ,\\
{\rm Det}\,\gr{\gamma}&=&\gamma_{2}\,{\rm e}^{-2\Gamma t}\; ,
\eea
defining the sets of coefficients $\Sigma_{i}$, $\alpha_{i}$, $\beta_{i}$,
$\gamma_{i}$. One has
%\begin{widetext}
\bea
%\begin{split}
{~}&
\hspace*{-3.2cm}
\Sigma_{4}=&\hspace*{-1.4cm}a^2 b^2+\frac{a^2}{4\mu^2_{2}}+\frac{b^2}{4\mu^2_{1}}
-a^2 b\frac{\cosh2r_{2}}{\mu_{2}}-a b^2\frac{\cosh2r_{1}}{\mu_{1}}
+ab\frac{\cosh2r_{1}\cosh2r_{2}}{\mu_{1}\mu_{2}}
-a\frac{\cosh2r_{1}}{4\mu_{1}\mu_{2}^2}-b\frac{\cosh2r_{2}}{4\mu_{1}^2\mu_{2}}
\nonumber\\
&&\hspace*{-1.4cm}+(c_{1}^2+c_{2}^2)
\left(a\frac{\cosh2r_{2}}{2\mu_{2}}
+\frac{b\cosh2r_{1}}{2\mu_{1}}
-\frac{\cosh2r_{1}\cosh2r_{2}}{4\mu_{1}\mu_{2}}-
\frac{\sinh2r_{1}\sinh2r_{2}\cos2\varphi_{2}}{4\mu_{1}\mu_{2}}-ab\right)
\nonumber \\
&&\hspace*{-1.4cm}+(c_{1}^2-c_{2}^2)\left(a\frac{\sinh2r_{2}\cos2\varphi_{2}}{2\mu_{2}}
+b\frac{\sinh2r_{1}}{2\mu_{1}}
-\frac{\sinh2r_{1}\cosh2r_{2}}{4\mu_{1}\mu_{2}}
-\frac{\cosh2r_{1}\sinh2r_{2}\cos2\varphi_{2}}{4\mu_{1}\mu_{2}}\right)
\nonumber\\
&&\hspace*{-1.4cm}+c_{1}^2c_{2}^2+\frac{1}{16\mu_{1}^2 \mu_{2}^2} \; , 
%\end{split}
\eea
\bea
{~}&
\hspace*{-3.2cm}
\Sigma_{3}=&\hspace*{-1.4cm}-2\frac{a^2}{4\mu^2_{2}}-2\frac{b^2}{4\mu^2_{1}}
+a^2 b\frac{\cosh2r_{2}}{\mu_{2}}+a b^2\frac{\cosh2r_{1}}{\mu_{1}}
-2ab\frac{\cosh2r_{1}\cosh2r_{2}}{\mu_{1}\mu_{2}}
+3a\frac{\cosh2r_{1}}{4\mu_{1}\mu_{2}^2}+3b\frac{\cosh2r_{2}}{4\mu_{1}^2\mu_{2}}
\nonumber\\
&&\hspace*{-1.4cm}-(c_{1}^2-c_{2}^2)\left(a\frac{\sinh2r_{2}\cos2\varphi_{2}}{2\mu_{2}}
+b\frac{\sinh2r_{1}}{2\mu_{1}}
-2\frac{\sinh2r_{1}\cosh2r_{2}}{4\mu_{1}\mu_{2}}
-2\frac{\cosh2r_{1}\sinh2r_{2}\cos2\varphi_{2}}{4\mu_{1}\mu_{2}}\right)
\nonumber\\
&&\hspace*{-1.4cm}-(c_{1}^2+c_{2}^2)\left(a\frac{\cosh2r_{2}}{2\mu_{2}}
+\frac{b\cosh2r_{1}}{2\mu_{1}}
-2\frac{\cosh2r_{1}\cosh2r_{2}}{4\mu_{1}\mu_{2}}-2
\frac{\sinh2r_{1}\sinh2r_{2}\cos2\varphi_{2}}{4\mu_{1}\mu_{2}}\right)\nonumber\\
&&\hspace*{-1.4cm}-\frac{1}{4\mu_{1}^2 \mu_{2}^2} \, ,
\eea
\bea
{~}&
\hspace*{-3.2cm}
\Sigma_{2}=&\hspace*{-1.4cm}\frac{a^2}{4\mu^2_{2}}+\frac{b^2}{4\mu^2_{1}}
+ab\frac{\cosh2r_{1}\cosh2r_{2}}{\mu_{1}\mu_{2}}
-3a\frac{\cosh2r_{1}}{4\mu_{1}\mu_{2}^2}
-3b\frac{\cosh2r_{2}}{4\mu_{1}^2\mu_{2}}
\nonumber\\
&&\hspace*{-1.4cm}-(c_{1}^2+c_{2}^2)\left(
\frac{\cosh2r_{1}\cosh2r_{2}}{4\mu_{1}\mu_{2}}+
\frac{\sinh2r_{1}\sinh2r_{2}\cos2\varphi_{2}}{4\mu_{1}\mu_{2}}\right)
\nonumber \\
&&\hspace*{-1.4cm}-(c_{1}^2-c_{2}^2)\left(
\frac{\sinh2r_{1}\cosh2r_{2}}{4\mu_{1}\mu_{2}}
+\frac{\cosh2r_{1}\sinh2r_{2}\cos2\varphi_{2}}{4\mu_{1}\mu_{2}}\right)
+\frac{1}{16\mu_{1}^2 \mu_{2}^2} \; ,
\eea
\be
\Sigma_{1}=
+a\frac{\cosh2r_{1}}{4\mu_{1}\mu_{2}^2}+b\frac{\cosh2r_{2}}{4\mu_{1}^2\mu_{2}}
-\frac{1}{4\mu_{1}^2 \mu_{2}^2} \; , 
\ee
\be
\Sigma_{0}=\frac{1}{16\mu_{1}^2\mu_{2}^2} \; ,
\ee
%\end{widetext}
\bea
\alpha_{2}&=&a^2-a\frac{\cosh2r_{1}}{\mu_{1}}+\frac{1}{4\mu_{1}^2}\; ,\\
\alpha_{1}&=&a\frac{\cosh2r_{1}}{\mu_{1}}-2\frac{1}{4\mu_{1}^2}\; ,\\
\alpha_{0}&=&\frac{1}{4\mu_{1}^2}\; , \\
\beta_{2}&=&b^2-b\frac{\cosh2r_{2}}{\mu_{2}}+\frac{1}{4\mu_{2}^2}\; ,\\
\beta_{1}&=&b\frac{\cosh2r_{2}}{\mu_{2}}-2\frac{1}{4\mu_{2}^2}\; ,\\
\beta_{0}&=&\frac{1}{4\mu_{2}^2}\; , \\
\gamma_{2}&=&c_{1}c_{2}\; .
\eea

The coefficients of Eq.~(\ref{dis4}), 
whose solution $k_{ent}$ allows to determine the entanglement time 
of an arbitrary two--mode Gaussian state, read
\bea
u&=&\Sigma_{4}\; ,\\
v&=&\Sigma_{3}\; ,\\
w&=&\Sigma_{2}-\alpha_{2}-\beta_{2}-|\gamma_{2}|\; ,\\
y&=&\Sigma_{1}-\alpha_{1}-\beta_{1}\; ,\\
z&=&\Sigma_{0}-\alpha_{0}-\beta_{0}+\frac14\; .
\eea 
%%%%%%%%%%%%%%%%%%%%%%%%%%%%%%%%%%%%%%%%%%%%%%%%


\begin{thebibliography}{99}

\bibitem{pati} {\em Quantum Information Theory with Continuous
Variables}, S. L. Braunstein and A. K. Pati Eds. (Kluwer, Dordrecht,
2002), and references therein.

\bibitem{rmp} {S. L. Braunstein and P. van Loock, quant-ph/0410100; Rev. Mod. Phys., 
to be published.}

\bibitem{tele} A. Furusawa, J. L. Sorensen, S. L. Braunstein, C. A.
Fuchs, H. J. Kimble, and E. S. Polzik, Science {\bf 282}, 706 (1998);
T. C. Zhang, K. W. Goh, C. W. Chou, P. Lodahl, and H. J. Kimble,
Phys. Rev. A {\bf 67}, 033802 (2003).

\bibitem{gran} F. Grosshans and P. Grangier, Phys. Rev. Lett. {\bf 88},
057902 (2002); F. Grosshans, G. Van Assche, J. Wenger,
R. Brouri, N. J. Cerf, and P. Grangier, Nature {\bf 421}, 238 (2003).

\bibitem{leggett83}A.\ O.\ Caldeira and A.\ J.\ Leggett, Physica A {\bf 121}, 587 (1983). 

\bibitem{zurek91}W.\ H.\ Zurek, Physics Today {\bf 44} (10), 36 (1991).

\bibitem{barnett} S. M. Barnett and P. M. Radmore,
{\em Methods in Theoretical Quantum Optics} (Clarendon Press, Oxford, 1997).

\bibitem{lee} C. T. Lee, Phys. Rev. A {\bf 44}, R2275 (1991).

\bibitem{kim02}M. S. Kim, W. Son, V. Buzek, and P. L. Knight, Phys.
Rev. A {\bf 65}, 032323 (2002).  

\bibitem{takeoka}M.\ Takeoka, M.\ Ban, and M.\ Sasaki, J.\ Opt.\ B: 
Quantum Semiclass.\ Opt.\ {\bf 4}, 114 (2002).

\bibitem{benedict}M. G. Benedict and A. Czirj\'ak, Phys. Rev. A {\bf 60}, 4034 (1999);
A. Kenfack and K. \.Zyczkowski, J. Opt. B: Quantum Semiclass. Opt.{\bf 6}, 396 (2004).

\bibitem{simon87} R. Simon, E. C. G. Sudarshan, and N. Mukunda,
Phys. Rev. A {\bf 36}, 3868 (1987).

\bibitem{prama}
    Arvind, B.\ Dutta, N.\ Mukunda, and R.\ Simon, Pramana {\bf 45}, 471
    (1995);
    quant-ph/9509002.

\bibitem{williamson}
    J.\ Williamson, Am.\ J.\ Math.\ {\bf 58}, 141 (1936); see also
    V.I.\ Arnold, {\em Mathematical Methods of Classical Mechanics},
    (Springer-Verlag, New York,
    1978) and R.\ Simon, S.\ Chaturvedi, and V.\ Srinivasan.,
    J.\ Math.\ Phys.\ {\bf 40}, 3632 (1999).

\bibitem{simon94} R. Simon, N. Mukunda, and B. Dutta, Phys. Rev. A {\bf 49}, 1567 (1994).

\bibitem{adesso} G. Adesso, A. Serafini, and F. Illuminati, Phys. Rev. Lett. {\bf 92}, 087901
(2004).

\bibitem{adesso2} G. Adesso, A. Serafini, and F. Illuminati,
Phys. Rev. A {\bf 70}, 022318 (2004).

\bibitem{adesso3} G. Adesso, A. Serafini, and F. Illuminati, 
Phys. Rev. Lett. {\bf 93}, 220504 (2004).

\bibitem{unitaryloc} A. Serafini, G. Adesso, and F. Illuminati, quant-ph/0411109,
and Phys. Rev. A {\bf 71} (2005), in press.

\bibitem{adesso4} G. Adesso and F. Illuminati, quant-ph/0410050.

\bibitem{ekefil} A. K. Ekert, C. Moura Alves, and
D. K. L. Oi, Phys. Rev. Lett. {\bf 88}, 217901 (2002);
R. Filip, Phys. Rev. A {\bf 65}, 062320 (2002).

\bibitem{fiura}J. Fiur\'a\v{s}ek and N. J. Cerf, Phys. Rev. Lett. {\bf 93}, 063601 (2004);
J. Wenger, J. Fiur\'a\v{s}ek, R.
Tualle-Brouri, N. J. Cerf, and Ph. Grangier, Phys. Rev. A {\bf 70}, 053812 (2004).

\bibitem{holevo99} A. S. Holevo, M. Sohma, and O. Hirota, Phys. Rev. A {\bf 59}, 1820 (1999).

\bibitem{serafozzi} A. Serafini, F. Illuminati, and S. De Siena, J. Phys. B: At. Mol. Op.
Phys. {\bf 37}, L21 (2004).

\bibitem{vedral01} L. Henderson and V. Vedral,
J. Phys. A:\ Math.\ Gen.\ {\bf 34}, 6899 (2001).

\bibitem{simon00} R. Simon, Phys. Rev. Lett. {\bf 84},
2726 (2000).

\bibitem{duan00}L.-M. Duan, G. Giedke, J. I. Cirac, and
P. Zoller, Phys. Rev. Lett. {\bf 84}, 2722 (2000).

\bibitem{vidwer} G. Vidal and R. F. Werner, Phys. Rev. A {\bf 65}, 032314 (2002).

\bibitem{eiserth} J. Eisert, PhD thesis, University of Potsdam (Potsdam, 2001).

\bibitem{auden03} K. Audenaert, M. B. Plenio, and J. Eisert, Phys. Rev. Lett. {\bf 90},
027901 (2003).

\bibitem{walls} D. Walls and G. Milburn, {\em Quantum optics}
(Springer Verlag, Berlin, 1994).

\bibitem{sqbath1} Possible squeezed reservoirs are treated in
M.-A. Dupertuis and S. Stenholm, J. Opt. Soc. Am. B {\bf 4},
1094 (1987); M.-A. Dupertuis, S. M. Barnett, and S. Stenholm, ibid. {\bf 4},
1102 (1987); Z. Ficek and P. D. Drummond, Phys. Rev. A {\bf 43}, 6247 (1991);
K. S. Grewal, Phys Rev A {\bf 67}, 022107 (2003);
see also C. W. Gardiner and P. Zoller, {\em Quantum Noise}
(Springer Verlag, Berlin, 1999),
M. S. Kim and N. Imoto, Phys. Rev. A {\bf 52}, 2401 (1995) and Ref.~\cite{walls}.

\bibitem{tomvit} P. Tombesi and D. Vitali,
Phys. Rev. A {\bf 50}, 4253 (1994); P. Tombesi, and D. Vitali,
Appl. Phys. B {\bf 60}, S69 (1995).

\bibitem{cir}J.\ F.\ Poyatos, J.\ I.\ Cirac, and P.\ Zoller,
Phys.\ Rev.\ Lett.\ {\bf 77}, 4728 (1996);
N. L\"utkenhaus, J. I. Cirac, and P. Zoller, Phys. Rev. A {\bf 57},
548 (1998).

\bibitem{wise}H. M. Wiseman and G. J. Milburn, Phys. Rev. Lett. {\bf 70}, 548 (1993);
H. M. Wiseman and G. J. Milburn, Phys. Rev. A {\bf 49}, 1350 (1994).

\bibitem{paris03} M. G. A. Paris, F. Illuminati, A. Serafini, and S. De Siena, Phys. Rev. A
{\bf 68}, 012314 (2003).

\bibitem{book} J.I.\ Cirac, J.\ Eisert, G.\ Giedke,  M.\
    Lewenstein, M.B.\ Plenio,
    R.F.\ Werner, and M.M.\ Wolf,
    %{\em Quantum information over systems with canonical coordinates},
    textbook in preparation (2004).

\bibitem{canali} A. Serafini, J. Eisert, and M. M. Wolf, 
Phys. Rev. A {\bf 71}, 012320 (2005).

\bibitem{brodier}O.\ Brodier and A.\ M.\ Ozorio de Almeida, Phys.\ Rev.\ E {\bf 69},
016204 (2004).

\bibitem{marian} P. Marian and T. A. Marian,
Phys. Rev. A {\bf 47}, 4487 (1993).

\bibitem{rajapla} R.\ W.\ Rendell and A.\ K.\ Rajagopal, 
Phys.\ Lett.\ A {\bf 279}, 175 (2001).

\bibitem{lloyd}V. Giovannetti, S. Guha, S. Lloyd, L. Maccone, and J. H. Shapiro,
Phys. Rev. A {\bf 70}, 032315 (2004); 
V. Giovannetti, S. Lloyd, L. Maccone, J. H. Shapiro, and B. J. Yen,
Phys. Rev. A {\bf 70}, 022328 (2004).

\bibitem{zurek93} W.\ H.\ Zurek, S.\ Habib, and J.\ P.\ Paz,
Phys.\ Rev.\ Lett. {\bf 70}, 1187 (1993).

\bibitem{venugopalan}A.\ Venugopalan, Phys.\ Rev.\ A {\bf 61}, 012102 (1999).

\bibitem{schroedinger} E.~Schr\"odinger, Naturwissenschaften {\bf 23}, 812 (1935).

\bibitem{yusto} B.~Yurke and D.~Stoler, Phys.~Rev.~Lett. {\bf 57}, 13 (1986).

\bibitem{catgene}A.~Mecozzi and P.~Tombesi, Phys.~Rev.~Lett. {\bf 58}, 1055 (1987);
B.~Yurke, W.~Schleich, and D.~F.~Walls, Phys.~Rev. A {\bf 42},
1703 (1990);
T.~Ogawa, M~Ueda, N.~Imoto, Phys.~Rev. A {\bf 43}, 6458 (1991);
M. Brune, S. Haroche, J. M. Raimond, L. Davidovich, and
N. Zagury, Phys. Rev. A {\bf 45}, 5193 (1992);
M.~Dakna, T.~Anhut, T.~Opatrny, L.~Kn\"oll, and D.~G.~Welsch,
  Phys.~Rev. A {\bf 55}, 3184 (1997);
S. Olivares, M. G. A. Paris and A. R. Rossi, Phys. Lett. A  {\bf 319}, 32 (2003);
A. R. Rossi, S. Olivares, and M. G. A. Paris, J. Mod. Opt. {\bf 51}, 1057 (2004);
M.~Paternostro, M.~S.~Kim, and B.~S.~Ham, Phys.~Rev. A {\bf 67},
  023811 (2003);
see also V. V. Dodonov, J. Opt. B: Quantum Semiclass. Opt. {\bf 4}, R1 (2002) and references therein.

\bibitem{haroche96}{M. Brune, E. Hagley, J. Dreyer, X. Ma\^{\i}tre,
A. Maali, C. Wunderlich, J.M. Raimond, and S. Haroche, Phys. Rev. Lett. {\bf 77}, 4887 (1996);
A.\ Auffeves, P.\ Maioli, T.\ Meunier, S.\ Gleyzes, G.\ Nogues, M.\ Brune, J.\ M.\ Raimond,
and S.\ Haroche,
Phys.\ Rev.\ Lett.\ {\bf 91}, 230405 (2003).}

\bibitem{monroecat}C.\ Monroe, D.\ M.\ Meekhof, B.\ E.\ King, and D.\ J.\ Wineland,
Science {\bf 272}, 1131 (1996).

\bibitem{myatt}C.\ J.\ Myatt, B.\ E.\ King, Q.\ A.\ Turchette, C.\ A.\ Sackett, D.\ Kielpinski,
W.\ M.\ Itano, C.\ Monroe, and D.\ J. Wineland, Nature {\bf 403}, 269 (2000).

\bibitem{walls85}D. F. Walls, G. J. Milburn, Phys. Rev. A {\bf 31}, 2403 (1985).

\bibitem{kennedy}T. A. B. Kennedy and D. F. Walls, Phys. Rev. A {\bf 37}, 152 (1988).

\bibitem{vitali}P. Goetsch, P. Tombesi, and D. Vitali, Phys. Rev. A {\bf 54}, 4519 (1996);
 S. Zippilli, D. Vitali, P. Tombesi, and J.-M. Raimond, ibid. {\bf 67},
052101 (2003).

\bibitem{el-ora}F. A. A. El-Orany, Phys. Rev. A {\bf 65}, 043814 (2002).

\bibitem{gatti}A. Serafini, S. De Siena, F. Illuminati, and M. G. A. Paris, J. Opt. B:
Quantum and Semiclass. Opt. {\bf 6}, S591 (2004).

\bibitem{garraway} B.\ M.\ Garraway, P.\ L.\ Knight, and M.\ B.\ Plenio,
Phys. Scr. {\bf T76}, 152 (1998).

\bibitem{crypto} H.-K. Lo and H. F. Chau, Science {\bf 283}, 2050 (1999);
  H. Zbinden, N. Gisin, B. Huttner, and W. Tittel, J. Cryptol. {\bf 13}, 207 (2000);
  T. Jennewein, C. Simon, G. Weihs, H. Weinfurter, and A. Zeilinger, Phys. Rev. Lett. {\bf 84},
  4729 (2000).

\bibitem{secom} K. M. Gheri, C. Saavedra, P. T\"orm\"a, J. I. Cirac, and P. Zoller, Phys. Rev. A
  {\bf 58}, R2627 (1998); S. J. van Enk, J. I. Cirac, and P. Zoller, Science {\bf 279}, 205 (1998).

\bibitem{non} S. Y. Kilin and D. B. Horosko, Phys. Rev. Lett. {\bf
74}, 5206 (1995); W. Leonski, S. Dyrting, and R. Tanas, J. Mod.  Opt.
{\bf 44}, 2105 (1997); A. Vidiella-Barranco, and J.  A. Roversi,
Phys. Rev. A {\bf 58 }, 3349 (1998); W. Leonski, Phys. Rev. A {\bf
54}, 3369 (1969).

\bibitem{con} M. G. A. Paris, Int. J. Mod. Phys. B {\bf 11}, 1913
(1997); M. Dakna, T. Anhut, T. Opatrny, L. Knoll, and D. G. Welsch,
Phys. Rev. A {\bf 55}, 3184 (1997); O. Steuernagel, Opt Comm {\bf 138} 
71 (1997).

\bibitem{fockfil} G. M. D'Ariano, L. Maccone, M. G. A. Paris, M. F. Sacchi, 
Phys. Rev. A {\bf 61} 053817 (2000); Fort. Phys. {\bf 48}, 511 (2000). 

\bibitem{eng} K. Vogel, V. M. Akulin, and W. P. Scheich,
Phys. Rev. Lett.  {\bf 71}, 1816 (1993).

\bibitem{serra} C. J. Villas-B\^oas,
  F. R. de Paula, R. M. Serra, and M. H. Y. Moussa, Phys. Rev. A {\bf 68}, 053808 (2003).

\bibitem{walther} B. T. H. Varcoe, S. Brattke, M. Weidinger, and H. Walther, 
  Nature {\bf 403}, 743 (2000); S. Brattke, B. T. H. Varcoe, and H. Walther, Phys. Rev. Lett. 
  {\bf 86} 3534 (2001).

\bibitem{brown2003} K. R. Brown, K. M. Dani, D. M. Stamper-Kurn, and 
  K. B. Whaley, Phys. Rev. A {\bf 67}, 043818 (2003).

\bibitem{kurizki} G. Harel and G. Kurizki, Phys. Rev. A {\bf 54}, 5410 (1996);
  G. Harel, G. Kurizki, J. K. McIver, and E. Coutsias, Phys. Rev. A {\bf 53}, 4534 (1996).

\bibitem{wineland} D. M. Meekhof, C. Monroe, B. E. King, W. M. Itano, and D. J. Wineland, 
Phys. Rev. Lett. 76, 1796 (1996).

\bibitem{nayak} N. Nayak, quant-ph/0308077.  

\bibitem{num}A. Serafini, S. De Siena, and F. Illuminati, Mod. Phys. Lett. B
{\bf 18}, 687 (2004).

\bibitem{tafeln} I. Gradstein and I. Ryshik, {\em Summen-, Produkt- und Integral- Tafeln}, 
  (Harri Deutsch Verlag, Thun $\cdot$ Frankfurt/M, 1981).
    
\bibitem{janszkim}J.\ Janszky, M.\ G.\ Kim, and M.\ S.\ Kim, Phys. Rev. A {\bf 53}, 502
(1996).

\bibitem{harochelh} S. Haroche, {\em Introduction to quantum optics and decoherence},
lectures held at Les Houches Summer School {\em Quantum entanglement and information 
processing} (2003). 

\bibitem{duan97} L.-M. Duan and G.-C. Guo, 
Quantum Semiclass. Opt. {\bf 9}, 953 (1997).

\bibitem{rajapra} A.\ K.\ Rajagopal and R.\ W.\ Rendell, Phys.\ Rev.\ A {\bf 63}, 022116 (2001).

\bibitem{hiroshima01} T. Hiroshima, Phys. Rev. A {\bf 63}, 022305 (2001).

\bibitem{scheel01} S. Scheel and D.-G. Welsch, 
Phys. Rev. A {\bf 64}, 063811 (2001).

\bibitem{paris02} M. G. A Paris {\em Entangled light and applications}
in {\em Trends in Quantum Physics}, V. Krasnoholovets and
F. Columbis Eds., p. 89 (Nova Publisher, Hauppauge NY, 2004).


\bibitem{kim03} D. Wilson, J. Lee, and M. S. Kim, J. Mod. Opt. {\bf 50}, 1809
(2003).

\bibitem{prauz03} J. S. Prauzner--Bechcicki, J.\ Phys.\ A:\ Math.\ Gen.\ {\bf 37}, L173 (2004).

\bibitem{serafini04} A. Serafini, F. Illuminati, M. G. A. Paris, and S. De Siena,
Phys. Rev. A {\bf 69}, 022318 (2004).

\bibitem{2max} S. M. Barnett, S. J. D. Phoenix,  Phys. Rev. A {\bf 40}, 2404
(1989); S. M. Barnett, S. J. D. Phoenix,  ibid. {\bf 44}, 535
(1991); M. G. A. Paris, ibid. {\bf 59}, 1615 (1999).

\bibitem{wieman99}C. E. Wieman, D. E. Pritchard, and D. J. Wineland
Rev. Mod. Phys. {\bf 71}, S253-S262 (1999).

\bibitem{vandersypen04}L. M. K. Vandersypen and I. L. Chuang
Rev. Mod. Phys. {\bf 76}, 1037 (2004).  

\bibitem{braunstein98}S. L. Braunstein and H. J. Kimble, Phys. Rev. Lett. {\bf 80}, 869 (1998).

\bibitem{adillu2} G. Adesso and F. Illuminati, quant-ph/0412125.

\end{thebibliography}
\end{document}